\documentclass{article}
\usepackage{color,graphicx,url,bookmark}
\def\Address#1{}
\def\Abstract#1{}
\def\Keywords#1{}
\def\proglang#1{\texttt{#1}}

\usepackage[english]{babel}
\usepackage[T1]{fontenc}

\usepackage{doxygen_for_JSS}

\usepackage{lmodern}
\usepackage{amsfonts}

\usepackage{fullpage}
\usepackage{amsmath,amssymb,amsthm}
\usepackage{mymath,maththm}

\usepackage{algpseudocode}

\usepackage{varioref}

\long\def\ammark#1{}

\renewcommand\P{{\mathbb P}}

\newcommand\A{{\mathcal A}} 
\newcommand\calE{{\mathcal E}} 
\newcommand\F{{\mathcal F}} 

\newcommand\B{{\overline B}}
\newcommand\X{{\overline X}}
\newcommand\Y{{\overline Y}}
\newcommand\Z{{\overline Z}}

\usepackage{xspace}

\def\ns{\textsl{ns}\xspace}

\title{Bit Recycling for Scaling Random Number Generators}

\author{Andrea C. G. Mennucci
\thanks{Andrea C. G. Mennucci,Scuola Normale Superiore, Pisa, Italy
    \texttt{andrea.mennucci@sns.it}}
}

\begin{document}

\maketitle

\begin{abstract}
  Many random number generators (RNG) are available nowadays; they are
  divided in two categories, \emph{hardware RNG}, that provide
  ``true'' random numbers, and \emph{algorithmic RNG}, also known as
  pseudo random number generators (PRNG). Both types usually generate random
  numbers $(X_n)_n$ as independent uniform samples in a range
  $0,\ldots 2^b-1$, with $b=8,16,32$ or $b=64$.  In applications, it
  is instead sometimes desirable to draw random numbers as independent
  uniform samples $(Y_n)_n$ in a range $1,\ldots M$, where moreover
  $M$ may change between drawings. Transforming the sequence $(X_n)_n$
  to $(Y_n)_n$ is sometimes known as \emph{scaling}.  We discuss
  different methods for scaling the RNG, both in term of mathematical
  efficiency and of computational speed.
\end{abstract}

\textbf{Keywords}: \emph{random number generation, uniform distribution, computational speed, entropy}

\textbf{MSC} 60-04

\section{Introduction}

We consider the following problem. We want to generate a sequence of
random numbers $(Y_n)_n$ with a specified probability distribution,
using as input a sequence of random numbers $(X_n)_n$ uniformly
distributed in a given range. There are various methods available;
these methods involve transforming the input in some way; for this
reason, these methods work equally well in transforming both
pseudo-random and true random numbers. One such method, called the
acceptance-rejection method, involves designing a specific algorithm,
that pulls random numbers, transforms them using a specific function,
tests whether the result satisfies a condition: if it is, the value is
accepted; otherwise, the value is rejected and the algorithm tries
again.

This kind of method has a defect, though: if not carefully
implemented, it throws away many inputs. Let's see a concrete example.
We suppose that we are given a random number generator
(RNG) that produces random bits, evenly
distributed, and independent\footnote{For example, repeatedly tossing
  a coin with the faces labeled $0,1$.}.  We want to produce a random
number $R$ in the range $\{1,2,3\}$, uniformly distributed and
independent.  Consider the following method.
\begin{Example} We draw two random bits; if the sequence is $11$, we
  throw it away and draw two bits again; otherwise we return the
  sequence as $R$, mapping $00, 01, 10$ to $1,2,3$.
\end{Example}
This is rather wasteful! The entropy in the returned random $R$ is
$\log_2(3)=1.585$bits; 
there is a $1/4$ probability that we throw away the input, so the
expected number of (pair of tosses) is $4/3$ and then expected number
of input bits is $8/3$; all together we are effectively using only
\[\frac{\log_2(3)}{8/3} = 59\%\]
of the input. 

The waste may be consider as an unnecessary slowdown of the RNG: if
the RNG can generate 1 bit in $1\mu s$, then, after the example
procedure, the rate has decreased to  $1.6 \mu s$ per
bit. Since a lot of effort was put in designing fast RNG in the near
past, then slowing down the rate by $+60\%$ is simply unacceptable.

Another problem in the example method above is that, although it is quite
unlikely, we can have a very long run of "11" bits.  This means
that we cannot guarantee that the above procedure will generate the
next number in a predetermined amount of time.

There are of course better solutions, as this \emph{ad hoc method}.
\begin{Example}(From \cite{DJ})~%
  \footnote{Actually, we had independently discovered the method
    here described; after writing this paper, though, we found
    it described in \cite{DJ}; \cite{DJ} is not a scientific publication,
    but rather a post in math blog, anyway it deserves credit.}
  We draw eight random bits, and consider
  them as a number $x$ in the range $0\ldots 255$; if the number is
  more than $3^5-1=242$, we throw it away and draw eight bits again;
  otherwise we write $x$ as five digits in base 3 and return these digits as
  5 random samples.
\end{Example}
This is much more efficient! The entropy in the returned five samples is
$5 \log_2(3)=7.92$bits; there is a $13/256$ probability that we throw away the input, so the
expected number of 8-tuples of inputs is $256/253$ and then expected number
of input bits is $2048/253$; all together we are now using
\[\frac{5 \log_2(3)}{2048/253} = 97\%\]
of the input.

In this paper we will provide a mathematical proof (section 2), and
discuss some method (section 3), to optimize the scaling of a RNG.
Eventually we 
discuss their efficiency, and present numerical speed tests.

\begin{Remark}
  A different approach may be to use a decompressing
  algorithm. Indeed, e.g., the \emph{arithmetic encoder} decoding
  algorithm, can be rewritten to decode a stream of bits to an output
  of symbols with prescribed probability distributions~\footnote{If
    interested, I have the code somewhere in the closet}.
  Unfortunately, it is quite difficult to mathematically prove that
  such an approach really does transform a stream of independent
  equidistributed bits into an output of independent random variables.
  Also, the \emph{arithmetic encoder} is complex, and this complexity
  would slow down the RNG, defeating one of the goals.  (Moreover, the
  \emph{arithmetic encoder} was originally patented.)
\end{Remark}

A note on notations. In all of the paper,
\ns is a \emph{nanosecond}, that is $10^{-9}$seconds. 
When $x$ is a real number, $\lfloor x \rfloor=\mathtt{floor}(x)$
is the largest integer that is less or equal than $x$.

\section{Process splitting}

Let $\natural=\{0,1,2,3,4,5\ldots\}$ be the set of natural numbers.

Let $(\Omega,\A,\P)$ a probability space,
let $(E,\calE)$ be a measurable space, and $\X$
a process  of i.i.d. random variables  $(X_n)_{n\in\natural}$ defined on  $(\Omega,\A,\P)$
and each taking values in  $(E,\calE)$.
We fix an event $S\in \calE$ such that
$\P\{X_i\in S\}\neq 0,1$; we define  $p_S\defeq\P\{X_i\in S\}$. 

We define a formal method of process splitting/unsplitting.

The \textbf{splitting} of $\X$ is the operation that generates three
processes $\B,\Y,\Z$, where $\B=(B_n)_{n\in\natural}$ is an
i.i.d.\ Bernoulli process with parameter $p_S$, and
$\Y=(Y_n)_{n\in\natural}$ and $\Z=(Z_n)_{n\in\natural}$ are processes
taking values respectively in $S$ and $E\setminus S$. The
\textbf{unsplitting} is the opposite operation.
These operations can be algorithmically and intuitively described by
the pseudocode in Figure~\vref{fig:s_us} (where processes are thought of as
\emph{queues of random variables}).

\begin{figure}[t!]
\noindent
\framebox{\noindent
  \begin{minipage}[t]{0.49\linewidth}
    {\centerline{$\X\mapsto (\B,\Y,\Z)$}}
     \hrule \smallskip
    \begin{algorithmic}[0]
      \Procedure{Splitting}{$\X$}
     \State initialize three empty queues $\B,\Y,\Z$
     \Repeat
     \State pop  X from $\X$
     \If {$X \in S$}
     \State push 1 onto $\B$ 
     \State push X onto $\Y$ 
    \Else
    \State push 0 onto $\B$ 
    \State push X onto $\Z$
    \EndIf
    \Until{forever}
    \EndProcedure
  \end{algorithmic}
\end{minipage}
\vline
  \begin{minipage}[t]{0.49\linewidth}
     {\centerline{$(\B,\Y,\Z)\mapsto \X$}}
     \hrule \smallskip
    \begin{algorithmic}[0]
      \Procedure{Unsplitting}{$(\B,\Y,\Z)$}
      \State initialize the  empty queue $\X$
     \Repeat
     \State pop  B from $\B$
     \If {$B = 1 $}
     \State pop Y from $\Y$ 
     \State push Y onto $\X$ 
    \Else
    \State pop Z from $\Z$ 
    \State push Z onto $\X$
    \EndIf
    \Until{forever}
    \EndProcedure
  \end{algorithmic}
\end{minipage}
}
\caption{Splitting and unsplitting.} \label{fig:s_us}
\end{figure}

The fact that the \emph{splitting} operation is invertible implies
that no entropy is lost when splitting. We will next show a very
important property, namely, that the splitting operation preserves
probabilistic independence.

\subsection{Mathematical formulation}
We now rewrite the above idea in a purely mathematical formulation.

We define the  Bernoulli process $(B_n)_{n\in\natural}$ by
\begin{equation}
  \label{eq:B}
  B_n\defeq
  \begin{cases}
    1 & X_n\in S\\
    0 & X_n\not\in S
  \end{cases}
\end{equation}
and the \textbf{times of return to success} as
\begin{eqnarray}
  \label{eq:ts}
  U_0&=&\min\{k: k\ge 0, B_k=1 \}\\
  U_n&=&\min\{k: k\ge 1+ U_{n-1}, B_k=1 \}, ~~n\ge 1
\end{eqnarray}
whereas the \textbf{times of return to unsuccess} are
\begin{eqnarray}
  \label{eq:tu}
  V_0&=&\min\{k: k\ge 0, B_k=0 \}\\
  V_n&=&\min\{k: k\ge 1+U_{n-1}, B_k=0 \}, ~~n\ge 1
\end{eqnarray}
it is well known that $(U_n),(V_n)$ are (almost certainly) well defined and finite.

We eventually define the processes  $(Y_n)_{n\in\natural}$ and $(Z_n)_{n\in\natural}$ by
\begin{equation}
  \label{eq:ZY}
  Y_n=X_{U_n}\quad Z_n=X_{V_n}
\end{equation}

\begin{Theorem}\label{thm:splitting}
  Assume that  $\X$  is a process of i.i.d.\ random variables. Let $\mu$ be the law of 
  $X_1$. We prove what follows. 
  \begin{itemize}\item 
    The random variables $B_n,Y_n,Z_n$ are independent.
  \item The variables of the same type are identically distributed:
    the variables $B_n$ have parameter $\P\{B_n=1\}=p_S$; 
    the  variables    $Y_n$ have law $\mu(\cdot ~|~S)$;
    the  variables    $Z_n$ have law $\mu(\cdot ~|~E\setminus S)$.
  \end{itemize}
  \begin{proof}
    It is obvious that $\B$ is a Bernoulli process of independent variables with parameter
    $\P\{B_n=1\}=p_S$.

    Let $K,M\ge 1$ integers.
    Let $u_0<u_1<\ldots u_K$ and $v_0<v_1<\ldots v_M$ be integers, and consider the event
    \begin{equation}
      A\defeq\{ U_0=u_0,\ldots U_K=u_K, V_0=v_0,\ldots V_M=v_M \} \label{eq:A}
    \end{equation}
    If $A\neq \emptyset$ then
    \begin{equation}
     A=\{ B_0 = b_0,\ldots , B_N=b_N\}\label{eq:A_as_B}    
   \end{equation}

    where $N=\max\{u_K,v_M\}$ and $b_j\in\{0,1\}$ are suitably chosen. Indeed, supposing that
    $N=u_K>v_M$, then we use the success times, and set that $b_j=1$ iff $j=u_k$ for a $k\le K$; whereas 
    supposing that
    $N=v_M>u_K$, then we use the unsuccess times, and set that $b_j=0$ iff $j=v_m$ for a $m\le M$.

    Let $\F_{K,M}$ be the family of all above events $A$ defined as per equation \eqref{eq:A}, for
    different choices of $(u_i),(v_j)$; let 
    \[\F=\bigcup_{K,M\ge 1}\F_{K,M}~~;\]
    let $\A^\B\subset\A$ be the sigma algebra generated by the  process $\B$.

    The above equality \eqref{eq:A_as_B} proves that
    $\F$ is a \emph{base} for  $\A^\B$: it is 
    stable by finite intersection, and it generates the sigma algebra $\A^\B$.
    \ammark{(possibly, up to negligeable sets).}

    Consider again $K,M\ge 1$ integers, and events $F_i,G_j\in\calE$ for $i=0,\ldots K$, $j=0,\ldots M$,
    and the event
    \[C=\{ Y_0\in F_0,\ldots Y_K\in F_K,Z_0\in G_0,\ldots Z_M\in G_M\}\in \A~~~; \]
    let $A\in \F_{K,M}$ non empty;  we want to show that
    \[ \P(C~|~ A)=\P(C)\]
    this will prove that $(\Y,\Z)$ are independent of $\B$,
    by arbitriness of $(F_i),(G_j),K,M$ and since $\F$ is a base for $\A^\B$.

    We fix $(u_i),(v_j)$ and define $A$ as in equation \eqref{eq:A}; we let $N=\max\{u_K,v_M\}$ and
    define $(b_n)$ as explained after equation \eqref{eq:A_as_B}.
    By defining  
    \[S^1\defeq S, S^0\defeq E\setminus S\]
    for notation convenience,  we can write 
    equation \eqref{eq:A_as_B} as
    \[A=\{      X_0\in S^{b_0},\ldots X_N\in S^{b_N} \}~~.\]
    Let $E_0\ldots E_N\in \calE$  be defined by
    \[E_n \defeq    \begin{cases}
      F_k & \text{if } n=u_k \text{ for a } k\le K\\
      G_m & \text{if } n=v_m \text{ for a } m\le M\\
      E   & \text{else}
    \end{cases}
    \]
    then we compute
    \begin{eqnarray*}
      \P(C~|~ A) &=& 
      \P(\{X_{u_0}\in F_0,\ldots X_{u_K}\in F_K, X_{v_0}\in G_0,\ldots X_{v_M}\in G_M \}~|~\{
      X_0\in S^{b_0},\ldots X_N\in S^{b_N} \})=\\ 
      &=& 
      \frac{\P\{X_{u_0}\in F_0,\ldots X_{u_K}\in F_K, X_{v_0}\in G_0,\ldots X_{v_M}\in G_M ~,~
      X_0\in S^{b_0},\ldots X_N\in S^{b_N} \}}     
    {\P\{X_0\in S^{b_0},\ldots X_N\in S^{b_N} \}}=\\ 
      &=&
      \frac{\prod_{n=0}^N\P\{X_n\in S^{b_n}\cap E_n\} }  
       {\prod_{n=0}^N\P\{X_n\in S^{b_n}\} } 
       =\prod_{n=0}^N\P(X_n\in  E_n | X_n\in S^{b_n})=\\ 
      &=&\prod_{k=0}^K\mu(F_k~|~S^1) \prod_{m=0}^M\mu(G_m~|~S^0)~~;
    \end{eqnarray*}
    the last equality is due to the fact that:
    when $n=u_k$ then $b_n=1$, when $n=v_m$ then $b_n=0$, and for all other $n$
    we have $E_n=E$.
    Since the last term does not depend on $A$, that is, on  $(u_i),(v_j)$,
    we obtain that  $(\Y,\Z)$ are independent of $\B$.

    \ammark{Also note that
      $  \P\{X_0\in S^{b_0},\ldots X_N\in S^{b_N} \}=    {p_S^{\overline b}(1-p_S)^{(N-\overbar b)}}$
    where \\
    $\overline b=\sum_{j=0}^N b_j=
    \begin{cases}
      K  &\text{ if } N=u_K>v_M\\
      N-M&\text{ if }N=v_M>u_K
    \end{cases}
    $
  }
  
  The above equality then also shows that
  \[\P\{ Y_0\in F_0,\ldots Y_K\in F_K,Z_0\in G_0,\ldots Z_M\in G_M\}=
  \prod_{k=0}^K\mu(F_k~|~S) \prod_{m=0}^M\mu(G_m~|~S^c)\]
  and this implies that $\Y,\Z$ are processes of independent variables,
  distributed as in the thesis.
  By associativity of the independence, we conclude that 
  the random variables $B_n,Y_n,Z_n$ are independent.
  \end{proof}
\end{Theorem}

\section{Recycling in uniform random number generation}
\label{sec:acci}
We now restrict our attention to the generation of independent uniformly
distributed integer valued random variables.  We will say that
\emph{$R$ is a random variable of modulus $M$} when $R$ is uniformly
distributed in the range $0,\ldots (M-1)$.  We will present some
different algorithms that process as input a sequence of independent
bits, and output a sequence of independent random variables of modulus
$M$; we will call them \textbf{uniform random functions} (URF for
short).

We present an algorithm, that we had thought of, and then found
(different implementation, almost identical idea) in \cite{DJ}. We
present the latter implementation.

The following algorithm \texttt{Uniform random by bit recycling} in
Figure~\vref{urbr}, given $n$, will return a random variable of
modulus $n$; note that $n$ can change between different calls to the
algorithm.

\begin{figure}[t!]
  \begin{algorithmic}[1]
    \State initialize the static integer variables $m = 1$ and $r = 0$
    \Procedure{Uniform random by bit recycling}{n}
    \Repeat
    \While {  m < N } \Comment{fill in the state}
    \State r : = 2 * r +    NextBit();
    \State m : = 2 * m; \Comment{r is a random variable of modulus m}
    \EndWhile
    \State q := $\lfloor$ m / n $\rfloor$; \Comment{integer division, rounded down} 
    \If  { r < n * q } \label{if}
    \State d : = r mod n  \Comment{remainder, is a random variable of modulus n}\label{def_r_as_Y}
    \State r : =  $\lfloor$ r / n $\rfloor$    \Comment{quotient, is random variable of modulus q}
    \State m : = q
    \State \Return d
    \Else
    \State r : = r - n * q \Comment{r is still a random variable of  modulus m } 
    \State m : = m - n * q \Comment{the procedure loops back to line 3}\label{def_r_as_Z}
    \EndIf
    \Until{forever}
    \EndProcedure
  \end{algorithmic}
  \caption{Algorithm ``uniform random by bit recycling''.} \label{urbr}
\end{figure}

It uses two internal integer variables, m and r, which are not reset 
each time the algorithm is called (in \proglang{C}, you would declare them 
as "static").  Initially, $m = 1$ and $r = 0$.

The algorithm has an internal constant parameter
$N$, which is a large integer such that $2N$ can still be represented 
exactly in the computer.  We must have $n < N$.
\footnote{We will show in next section that it is best to have $n<<N$.}
The algorithm draws randomness from a function \texttt{NextBit()} that returns a 
random bit.

Here is an informal discussion of the algorithm, in the words of the
original author \cite{DJ}.  \emph{
  At line \ref{def_r_as_Y}, as r is between 0
  and $(n*q - 1)$, we can consider r as a random variable of modulus
  $n*q$.  As this is divisible by n, then $d:=(r \mod n)$ will be
  uniformly distributed, and the quotient $\lfloor r/n\rfloor$ will be uniformly
  distributed between 0 and $q - 1$. }

Note that the theoretical running time is unbounded; we will though
show in the next section that an accurate choice of parameters
practically cancels this problem.

\subsection{Mathematical analysis of the efficiency}

We recall this simple idea.
\begin{Lemma}\label{lemma:QD}
  Suppose $R$ is  a random   variable of modulus $M  N$;
  we perform the integer division
  $R= Q N + D$ where $Q\in\{ 0,\ldots (M-1) \}$ is the quotient and
  $D\in\{ 0,\ldots (N-1) \}$ is the remainder;
  then $Q$ is  a random   variable of modulus $M$ and $D$ is  a random   variable of modulus $N$;
  and $Q,D$ are independent.
\end{Lemma}

\begin{Proposition}
  Let us assume that repeated calls of \texttt{NextBit()} return a
  sequence of independent equidistributed bits.
  Then the above algorithm \texttt{Uniform random by bit recycling} in
  Figure~\vref{urbr} will return a sequence of independent
  and uniformly distributed numbers.

  \begin{proof}
    We sketch the proof.  We use the Lemma~\ref{lemma:QD} and
    the Theorem~\ref{thm:splitting}.
    Consider the notations in the second section. The bits returned by
    the call \texttt{NextBit()} builds up the process $\X$.
    When reaching the \texttt{if} (line \ref{if} in the pseudocode at
    page \pageref{urbr}), the choice $r< n * q$ is the choice of
    the value of $B_n$ in equation \eqref{eq:B}. This (virtually) builds the
    process $\B$.
    At line \ref{def_r_as_Y} $r$ is a variable in the process $\Y$;
    since it is of modulus $n q$, we return (using the lemma) the
    remainder as $d$, that is a random variable of modulus $n$, and
    push back the quotient into the state.
    At line \ref{def_r_as_Z} we would be defining a variable in the
    process $\Z$, that we push back into the state.
  \end{proof}
\end{Proposition}
The ``pushing back'' of most of the entropy back into the state
recycles the bits, and improves greatly the efficiency.

The only wasted bits are related to the fact that the algorithm is
throwing away the mathematical stream $\B$.  Theoretically, if this
stream would be feeded back into the state (for example, by employing
Shannon-Fano-Elias coding), then efficiency would be exactly
$100\%$.\footnote{But this would render difficult to prove that the
  output numbers are independent...}  Practically, the numbers $N$ and
$n$ can be designed so that this is totally unneeded.
\begin{Remark}
  Indeed, consider the implementation (see the code in the next sections)
  where the internal state is stored as 64bit unsigned integers,
  whereas $n$ is restricted to be 32bit unsigned integer; so the
  internal constant is $N=2^{62}$ while $n\in \{2\ldots 2^{32}-1\}$, When
  reaching the \texttt{if} at line \ref{if}, $m$ is in the range $2^{62}\le m <
  2^{64}$, and $r$ is uniform of modulus $m$; but $m-n*\lfloor m/n\rfloor$ 
  is less than $n$, that is, less than $2^{32}$; so the probability that $r
  \ge n * q$ at the \texttt{if} is less than $1/2^{30}$.
\end{Remark}
In  particular, this means that each $B_n$ in the mathematical stream
$\B$ contains $\sim 10^{-8}$bits of entropy, so there is no need to
recycle them.

Indeed, in the numerical experiments we found out that the
algorithm wastes $\sim 30$ input bits on a total of $\sim 10^9$ input
bits (!) this is comparable to the entropy of the internal state (and
may also be due to numerical error in adding up $\log_2()$ values).

Also, this choice of parameters ensures that the algorithm will never
practically loop twice before returning. 
When the condition in the \texttt{if} at line  \ref{if} is false,
we will count it as a \textbf{failure}.
 In $\sim 10^{10}$ calls to
the algorithm, we only experienced 3 failures.
\footnote{For this reason, the
  \texttt{else} block may be omitted with no big impact on the quality
  of the output -- we implement this idea in the algorithm
  \texttt{uniform\_random\_by\_bit\_recycling\_cheating}.}


\begin{figure}[t!]
  \begin{algorithmic}[1]
    \Procedure{Uniform random simple}{n}
    \Repeat
    \State r : = GetRandomBits(b); \Comment{fill the state with b bits}
    \State q := $\lfloor$ N / n $\rfloor$; \Comment{integer division, rounded down} 
    \If { r < n * q }\label{simple:if}
    \State \Return r mod n     \Comment{remainder, is random variable of modulus n}
    \EndIf \Comment{otherwise, start all over again}
    \Until{forever}
    \EndProcedure
  \end{algorithmic}
  \caption{``Uniform random simple'' algorithm;
    in our tests $N=2^b$ or $N=2^b-1$, whereas $b=32,40,48,64$.}\label{simple}
\end{figure}

\section{Speed, simple {vs} complex algorithms}

We now consider the algorithm \texttt{Uniform random simple} in
Figure~\vref{simple}.

Again, when the condition in the \texttt{if} at line  \ref{simple:if} is false,
we will count it as a \textbf{failure}.

This algorithm will always call the original RNG to obtain $b$ bits,
regardless of the value of $n$.  When the algorithm fails, 
it starts again and again draws $b$ bits.  This is inefficient
in terms of entropy: for small values of $n$ it will produce far
less entropy than it consume. But, will it be slower or faster than
our previous algorithm?  It turns out that the answer pretty much
depends on the speed of the back-end RNG (and this is unsurprising);
but also on how much time it takes to compute the basic operations
``integer multiplications'' $q*n$ and ``integer division'' $\lfloor
N/n \rfloor$: we will see that, in some cases, these operations are so
slow that they defeat the efficiency of the algorithm \texttt{Uniform
  random by bit recycling}.

\section{Code structure}

\subsection{Data flow}


This is the flow of random data;
details are in the following sections.
\begin{center}
  \includegraphics[width=0.5\linewidth]{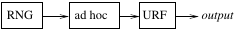}
\end{center}
Random bits (32bit or 64bit for each call) are generated by a backend RNG,
and then these bits are repacked, for convenience, by ad--hoc
functions, each one to return them in blocks of either 1,2,8 or 16 bits.  These
ad--hoc functions are used as input by some implementations of
\emph{uniform random functions}.

\subsection{Back-end PRNG}
\label{sec:back_end}
To test the speed of the following algorithms, we used four different back-end
pseudo random number generators (PRNG).
\begin{description}
\item[(sfmt\_sse)] The \emph{ SIMD oriented fast Mersenne Twister (SFMT)} ver. 1.3.3 by 
  Mutsuo Saito and Makoto Matsumoto (compiled with SSE support). See \cite{SFMT}.
\item[(xorshift)] The \emph{xorshift} generator by G. Marsaglia. See \cite{xorshift}.
\item[(sfmt\_sse\_md5)] As sfmt\_sse above, but moreover the output is cryptographically protected using
  the \texttt{MD5} algorithm.
\item[(bbs260)] The Blum-Blum-Shub algorithm, with two primes of size $\sim 130$bit. See \cite{bbs}.
\end{description}
The last two were home-made, as examples of slower but (possibly) cryptographically strong RNG
\footnote{The author makes no guarantees, though, that the implemented versions are really
  good and cryptographically strong RNGs --- we are interested only in their speeds.}.
All of the above were uniformized to implement two functions,
\texttt{my\_gen\_rand32()} and \texttt{my\_gen\_rand64()}, that
return (respectively) a 32bit or a 64bit unsigned integer, uniformly distributed.
The \proglang{C} code for all the above is in the appendix~\ref{sec:backend_C}.
The speeds of the different RNGs are listed in the tables in 
Sec.~\vref{sec:adhoc_speeds}.

We also prepared a simple \emph{counter} ``RNG'' algorithm, that
returns numbers that are in arithmetic progression; since it is very
simple, it is useful to assess the overhead complexity in the testing
code itself; this overhead is on the order of 2 to 
to 4 \ns, depending on the CPUs.

\subsection{{Ad hoc} functions}

We implemented some \emph{ad hoc} functions, that are
then used by the uniform RNGs (that are described in the next section).
\begin{description}
\item[NextBit] returns a bit
\item[Next2Bit] returns two bits
\item[NextByte] returns 8 bits
\item[NextWord] returns 16 bits
\end{description}
For any of the above, we prepared many variants,
that internally call either the \texttt{my\_gen\_rand32()} or \texttt{my\_gen\_rand64()}
calls (see the \proglang{C} code in Sec.~\vref{sec:adhoc_C})
and then we benchmarked them in each architecture,
to choose the faster one (that is then used by the URFs).
\footnote{We had to make an exception for when the back-end RNG
  is based on \texttt{SFMT}, since \texttt{SFMT} cannot mix 
  64bit and 32bit random number generations: in that case,
  we forcibly used the 32bit versions (that in most of our benchmarks
  are anyway slightly faster).
}

We also prepared a specific method (that is not used for the uniform RNGs):
\begin{description}
\item[NextCard] returns a number uniformly distributed in the range
  $0\ldots 51$ (it may be thought of as a card randomly drawn from a
  deck of cards).
\end{description}

The detailed timings are in the tables in Sec.~\vref{sec:adhoc_speeds}.

\subsection{Uniform RNGs}
We implemented nine different versions of uniform random generators. The C code is in 
Sec.~\vref{sec:ur_C};
we here briefly describe the ideas.  Four versions are based on the ``simple'' generator in Fig.~\ref{simple}:
{\small
\begin{description}
\item[uniform\_random\_simple32] uses 32bit variables internally, $N=2^{32}-1$, and consumes a 32bit random number, (a call to \texttt{my\_gen\_rand32()}) each time
\item[uniform\_random\_simple40] uses 64bit variables internally, $N=2^{40}$, and calls \texttt{my\_gen\_rand32()} and \texttt{NextByte()} each time
\item[uniform\_random\_simple48] uses 64bit variables internally, $N=2^{48}$, and calls \texttt{my\_gen\_rand32()} and \texttt{NextWord()}  each time
\item[uniform\_random\_simple64] uses 64bit variables internally, $N=2^{64}-1$, and calls \texttt{my\_gen\_rand64()}  each time.
\end{description}
}

Then there are three versions  based on the ``bit recycling'' generator in fig.~\ref{urbr} (all use 64bit variables internally):
{\small
\begin{description}
\item[uniform\_random\_by\_bit\_recycling] is the code in  fig.~\ref{urbr} (but it refills the state by popping two bits at a time)
\item[uniform\_random\_by\_bit\_recycling\_faster]  it refills the state by popping words, bytes
  and pairs of bits, for improved efficiency
\item[uniform\_random\_by\_bit\_recycling\_cheating] as the ``faster'' one, but the \emph{if/else} block is not  implemented,
  and the modulus $r\mod n$ is always returned; this is not mathematically exact, but the probability that it is inexact
  is $\sim 2^{-30}$.
\end{description}
}
Moreover there are ``mixed'' methods
{\small
\begin{description}
\item[uniform\_random\_simple\_recycler] uses 32bit variables internally, keeps an internal state that is sometimes initialized
  but not refilled each time (so, it is useful only for small $n$),
\item[uniform\_random\_by\_bit\_recycling\_32] when $n<2^{29}$, it implements the ``bit recycling'' code using 32bit variables; when
  $2^{29}\le n < 2^{32}$, it implements a ``simple''--like method, using only bit shifting.
\end{description}
}
We tested them in all of the architectures
(see Section~\vref{sec:arch}), for different values of 
the modulus $n$, and graphed the results (see appendix~\vref{sec:graph_speeds}).

\section{Numerical tests}

\subsection{Architectures}
\label{sec:arch}
The tests were performed in six different architectures,
\begin{description}
\item[(HW1)]\emph{Intel\textregistered{}  Core \texttrademark{} 2 Duo CPU     E7500    2.93GHz}, in i686 mode,
\item[(HW2)]\emph{Intel\textregistered{}  Core \texttrademark{} 2 Duo CPU     P7350    2.00GHz}, in i686 mode,
\item[(HW3)]\emph{AMD Athlon\texttrademark{}  64 X2 Dual Core Processor 4200+}, in x86\_64 mode,
\item[(HW4)]\emph{AMD Athlon  \texttrademark{}  64 X2 Dual Core Processor 4800+}, in x86\_64 mode,
\item[(HW5)]\emph{Intel\textregistered{}  Core \texttrademark{} 2 Duo CPU     P7350    2.00GHz}, in x86\_64 mode,
\item[(HW6)]\emph{Intel \textregistered{}  Xeon \textregistered{} CPU 5160 3.00GHz }, in x86\_64 mode.
\end{description}
In the first five cases, the host was running a \emph{Debian GNU/Linux}
or \emph{Ubuntu} O.S.\ , and the code was compiled using \emph{gcc
  4.4}, with the optimization flags \\
{\small \verb+ -march=native -O3 -finline-functions -fno-strict-aliasing -fomit-frame-pointer -DNDEBUG + }.\\
In the last case, the O.S.\ was \emph{Gentoo} and the  code was compiled using \emph{gcc 4.0}
with flags \\
{\small \verb+ -march=nocona -O3 -finline-functions -fno-strict-aliasing -fomit-frame-pointer -DNDEBUG +}.

\subsection{Timing}
To benchmark the algorithms, we computed the process time using both
the Posix call \texttt{clock()} (that returns an approximation of
processor time used by the program) and the CPU \texttt{TSC} (that
counts the number of CPU ticks). When benchmarking one of the
above back-end RNGs or the \emph{ad hoc} functions, we called it in
repeated loops of $2^{24}$ iterations each, repeating them for at
least 1 second of processor time;
and then compared the data provided by \texttt{TSC}
and \texttt{clock()}. We also prepared a statistics of the values
\[\mathtt{cycles\_per\_clock}:=\frac{\Delta \texttt{TSC}}{\Delta \texttt{clock()}}\]
so that we could convert CPU cycles to nanoseconds; we verified that
the standard deviation of the logarithm of the above quantity was
usually less than $1\%$.

To avoid over-optimization of the compiler,
the results of any benchmarked function was \emph{xor}-ed
in a \emph{bucket} variable, that was then printed on screen.

During benchmarking, we disabled the CPU power-saving features, forcing
the CPU to be at maximum performance (using the \texttt{cpufreq-set}
command) and also we tied the process to one core (using the
\texttt{taskset} command).

Unfortunately the \texttt{clock()} call, in GNU/Linux systems, has a
time resolution of $0.01sec$, so it was too coarse to be used for the
graphs in section \ref{sec:graph_speeds}: for those graphs, only the
\texttt{TSC} was used (and then cycles were converted to nanoseconds, using the average value of 
\texttt{cycles\_per\_clock}).

\section{Conclusions}

While efficiency is exactly mathematically assessed, computational speed 
is a more complex topic, and sometimes quite surprising. We report some considerations.
\begin{enumerate}
\item In our Intel\texttrademark{} 
  CPUs running in 32bit mode, integers divisions and remainder computation
  using 64bit variables are quite slow: in \texttt{HW1}, each such operations
  cost $\sim 15$\ns. 

\item Any bit recycling method that we could think of needs at least four
  arithmetic operations for each result it produces; moreover there is
  some code to refill the internal state.

\item 
  In the same CPUs, the cost of \emph{bit shifting} or \emph{xor} 
  operations are on the order of 3 \ns, even on 64bit variables;
  moreover the back-ends \texttt{sfmt\_sse} and \texttt{xorshift}
  can produce a 32bit random number in $\sim 5$ \ns.

\item So, unsurprisingly, when the back end is \texttt{sfmt\_sse} and
  \texttt{xorshift}, and the Intel\texttrademark{} CPU runs in 32bit
  mode, the fastest methods are the ``simple32'' and
  ``simple\_recycler'' methods, that run in $\sim 10$\ns; and
  the bit recycling methods are at least 5 times slower than those.

\item When the back-end is \texttt{sfmt\_sse} and \texttt{xorshift},
  but the the Intel\texttrademark{} CPU runs in 64bit mode, the
  fastest method are still the ``simple32'' and ``simple\_recycler''
  methods; the bit recycling methods are twice slower.

\item When the back-ends RNGs are \texttt{sfmt\_sse} or
  \texttt{xorshift}, in the AMD\texttrademark{} CPUs, the ``simple32''
  and ``simple\_recycler'' take $\sim 40$\ns; this is related
  to the fact that 32bit division and remainder computation need $\sim
  20$\ns (as is shown in
  sec.~\ref{sec:graph_arith_speeds}). So these methods are much slower
  than the back-ends RNGs, that return a 32bit random number in $\sim
  6$\ns.  

  One consequence is that, since the
  \texttt{uniform\_random\_by\_bit\_recycling\_32} for $n>2^{29}$ uses
  a ``simple''--like method with only bit shifting, then it is much
  faster than the ``simple32'' and ``simple\_recycler''.

  What we cannot explain is that, in the same
  architectures, the \texttt{NextCard32} function, that implements the
  same type of operations, runs in $\sim 8$\ns (!)

  (We also tried to test the above with different optimizations. Using
  the \texttt{xorshift} back-end, setting optimization flags to be
  just \texttt{-O0}, \texttt{NextCard32} function takes $\sim 21$\ns;
  setting it to \texttt{-O}, it takes $\sim 12$\ns.)

\item The back-end RNGs \texttt{sfmt\_sse\_md5} or \texttt{bbs260}, are
  instead much slower, that is, \texttt{sfmt\_sse\_md5} needs $\sim
  300$ \ns to produce a 32bit number, and \texttt{bbs260}
  needs $\sim 500$ \ns when the CPU runs in 64bit mode and
  more than a microsecond (!) in 32bit mode.

  In this case, the bit recycling methods are usually faster. Their
  speed is dominated by how many times the back-end RNGs is called, so
  it can be estimated in terms of \emph{entropy bitrate}, and indeed
  the graphs are (almost) linear (since the abscissa is in logarithmic scale).

\item 
  One of the biggest surprises comes from the \texttt{NextCard}
  functions. There are four implementations:
  \begin{itemize}\item 
    using 64bit or 32bit variables;
  \item computing a result for each call, or precomputing them and
    storing in an array (the ``prefilled'' versions).
  \end{itemize}
  The speed benchmarks give discordant results.  When using the faster
  back-ends \texttt{sfmt\_sse} and \texttt{xorshift}, the ``prefilled''
  versions are slower.  When using the slower back-ends
  \texttt{sfmt\_sse\_md5} or \texttt{bbs260}, the 64bit ``prefilled''
  version is the fastest in  Intel\texttrademark{} CPUs;
  but it is instead much slower than the ``non prefilled''
  version in AMD\texttrademark{} CPUs.
  It is possible that the cache misses are
  playing a r\^ole in this, but we cannot provide a good explanation.

\item Curiously, in some Intel\texttrademark{} CPUs, the time needed
  for an integer arithmetic operation depends also on the
  \emph{values} of the operands (and not only on the bit sizes of the
  variable)! See the graph in sec.~\ref{sec:graph_arith_speeds}. So,
  the speed of the functions depend on the value of the modulus  $n$. 
  This is the reason why some the graphs are all oscillating in
  nature.

  In particular, when we looked at the graphs for \emph{Core 2}
  architectures in 32bit mode, by looking at the graphs of the
  functions \texttt{simple\_40}, \texttt{simple\_48} (where
  $N=2^{40},2^{48}$ constant) we noted that the operations $q:=N/n,
  qn:=n*q$ are $\sim 10$ \ns slower when $n<N 2^{-32}$ than
  when $n > N 2^{-32}$. This is similar to what is seen
  in the graphs in sec.~\ref{sec:graph_arith_speeds}.

  Instead the speed graphs in AMD\texttrademark{} CPUs are almost linear,
  and this is well explained by the average number of needed operations.
\end{enumerate}

Summarizing, the speeds are quite difficult to predict; if a uniform
random generator is to be used for $n$ in a certain range, and the
back-end RNG takes approximatively as much time as 4 integer
operations in 64bits, then the only sure way to decide which algorithm
is the fastest one is by benchmarking.  If a a uniform random
generator is to be used for a constant and specific $n$ (such as in
the case of the \texttt{NextCard} function), there may be different
strategies to implement it, and again the only sure way to decide
which algorithm is the fastest one is by benchmarking.

\appendix

\section{Test results}

\subsection{Speed of back-end RNGs and {Ad hoc} functions}
\label{sec:adhoc_speeds}

These tables list the average time (in nanoseconds) of the back-end
RNGs and the \emph{ad hoc} functions (see the C code in
Sec.~\vref{sec:adhoc_C}); these same data are plotted as red crosses
in the plots of the next section.  The \emph{ad hoc} functions are
grouped in families (that are marked by the horizontal rules); for
each family and each hardware, the fastest function is marked blue;
competitors that differ less than $10\%$ are italic and blue;
competitors that are slower more than $50\%$ are red.

{\scriptsize\setlength{\tabcolsep}{2pt}
\noindent\begin{tabular}{l|rrrrrr|rrrrrr|}
 & \multicolumn{6}{|c|}{sfmt\_sse}   &  \multicolumn{6}{|c|}{xorshift } \\
 &  HW1  &  HW2  &  HW3  &  HW4  &  HW5  &  HW6  &  HW1  &  HW2  &  HW3  &  HW4  &  HW5  &  HW6 \\
\hline
Next2Bit32&{\color{blue}{2.6}}&{\color{blue}{3.9}}&{\color{blue}{4.9}}&{\color{blue}\emph{4.8}}&{\color{blue}\emph{3.9}}&{\color{blue}\emph{2.6}}&{\color{blue}{2.5}}&{\color{blue}{3.7}}&{\color{blue}\emph{4.7}}&{\color{blue}\emph{4.2}}&{\color{blue}{3.7}}&{2.8}   \\ 
Next2Bit64&{\color{red}{4.2}}&{4.7} &{\color{blue}\emph{5.0}}&{\color{blue}{4.4}}&{\color{blue}{3.7}}&{\color{blue}{2.5}}&{3.3} &{\color{red}{6.3}}&{\color{blue}{4.5}}&{\color{blue}{4.0}}&{\color{blue}{3.7}}&{\color{blue}{2.4}}  \\ 
\hline
NextBit32&{\color{blue}{2.5}}&{\color{blue}{3.7}}&{\color{blue}{4.6}}&{\color{blue}{4.2}}&{\color{blue}\emph{3.7}}&{\color{blue}\emph{2.8}}&{\color{blue}{2.5}}&{\color{blue}{3.6}}&{\color{blue}\emph{4.4}}&{\color{blue}\emph{3.9}}&{\color{blue}{3.6}}&{\color{blue}{2.4}}  \\ 
NextBit32\_by\_mask&{\color{blue}{2.5}}&{\color{blue}{3.7}}&{\color{blue}\emph{4.9}}&{\color{blue}\emph{4.4}}&{\color{blue}\emph{3.7}}&{\color{blue}\emph{2.8}}&{\color{blue}{2.5}}&{\color{blue}{3.6}}&{\color{blue}\emph{4.4}}&{\color{blue}\emph{3.9}}&{\color{blue}{3.6}}&{2.7}   \\ 
NextBit64&{3.2} &{4.6} &{\color{blue}\emph{4.7}}&{\color{blue}{4.2}}&{\color{blue}{3.6}}&{\color{blue}{2.8}}&{3.2} &{4.7} &{\color{blue}{4.3}}&{\color{blue}{3.8}}&{\color{blue}{3.6}}&{\color{blue}{2.4}}  \\ 
\hline
NextByte32&{3.5} &{5.1} &{\color{blue}{6.1}}&{\color{blue}{5.4}}&{5.1} &{3.8} &{\color{blue}\emph{3.0}}&{\color{blue}\emph{4.4}}&{\color{blue}{5.1}}&{\color{blue}{4.5}}&{\color{blue}\emph{4.5}}&{3.3}   \\ 
NextByte64&{\color{blue}\emph{3.3}}&{5.6} &{\color{blue}\emph{6.5}}&{\color{blue}\emph{5.8}}&{\color{blue}{4.4}}&{\color{blue}{3.1}}&{\color{blue}\emph{3.0}}&{\color{blue}\emph{4.4}}&{6.2} &{5.2} &{\color{blue}{4.3}}&{\color{blue}{2.8}}  \\ 
NextByte64\_prefilled&{\color{blue}{3.1}}&{\color{blue}{4.6}}&{\color{blue}\emph{6.3}}&{\color{blue}\emph{5.6}}&{\color{blue}\emph{4.5}}&{3.4} &{\color{blue}{2.9}}&{\color{blue}{4.3}}&{5.8} &{6.0} &{\color{blue}\emph{4.6}}&{\color{blue}\emph{3.0}}  \\ 
\hline
NextCard32&{\color{blue}{3.5}}&{\color{blue}{5.1}}&{8.0} &{7.1} &{\color{blue}{5.5}}&{7.0} &{\color{blue}{3.4}}&{\color{blue}{4.9}}&{\color{blue}{5.6}}&{\color{blue}{5.0}}&{\color{blue}{5.2}}&{6.9}   \\ 
NextCard32\_prefilled&{\color{red}{6.1}}&{\color{red}{9.0}}&{\color{red}{24.4}}&{\color{red}{21.7}}&{\color{red}{11.4}}&{7.1} &{\color{red}{6.0}}&{\color{red}{15.6}}&{\color{red}{22.2}}&{\color{red}{19.7}}&{\color{red}{11.1}}&{7.1}   \\ 
NextCard64&{\color{red}{22.0}}&{\color{red}{32.2}}&{\color{blue}{6.5}}&{\color{blue}{5.8}}&{7.0} &{\color{blue}{4.9}}&{\color{red}{22.4}}&{\color{red}{32.8}}&{6.6} &{5.8} &{7.0} &{\color{blue}{4.9}}  \\ 
NextCard64\_prefilled&{\color{red}{22.0}}&{\color{red}{32.2}}&{\color{red}{37.9}}&{\color{red}{33.5}}&{\color{red}{20.5}}&{\color{blue}\emph{5.4}}&{\color{red}{24.0}}&{\color{red}{37.0}}&{\color{red}{37.4}}&{\color{red}{33.1}}&{\color{red}{20.4}}&{\color{blue}\emph{5.3}}  \\ 
\hline
NextWord32&{\color{blue}{4.2}}&{\color{blue}{6.2}}&{7.5} &{6.7} &{6.5} &{4.5} &{\color{blue}{3.4}}&{\color{blue}{5.0}}&{\color{blue}{5.7}}&{\color{blue}{5.0}}&{\color{blue}\emph{5.0}}&{\color{blue}{3.3}}  \\ 
NextWord64&{\color{blue}{4.2}}&{\color{blue}\emph{6.5}}&{\color{blue}{6.5}}&{\color{blue}{5.8}}&{\color{blue}{5.2}}&{\color{blue}{3.7}}&{4.2} &{5.9} &{\color{blue}{5.7}}&{\color{blue}{5.0}}&{\color{blue}{4.9}}&{\color{blue}\emph{3.4}}  \\ 
\hline
my\_gen\_rand32&{\color{blue}{3.7}}&{\color{blue}{5.4}}&{\color{blue}{6.5}}&{\color{blue}{5.7}}&{\color{blue}{5.4}}&{\color{blue}{3.9}}&{\color{blue}{3.5}}&{\color{blue}{5.1}}&{\color{blue}{5.0}}&{\color{blue}{4.4}}&{\color{blue}{5.4}}&{\color{blue}{3.5}}  \\ 
my\_gen\_rand64&{4.2} &{6.3} &{8.4} &{7.4} &{6.3} &{5.2} &{4.5} &{6.9} &{6.8} &{6.0} &{7.4} &{4.8}   \\ 
\end{tabular}

\noindent\begin{tabular}{l|rrrrrr|rrrrrr|}
 & \multicolumn{6}{|c|}{sfmt\_sse\_md5} & \multicolumn{6}{|c|}{bbs260 } \\
 &  HW1& HW2& HW3& HW4& HW5& HW6& HW1& HW2& HW3& HW4& HW5& HW6 \\
\hline
Next2Bit32&{\color{red}{18.7}}&{\color{red}{27.5}}&{\color{red}{30.2}}&{\color{red}{26.7}}&{\color{red}{32.0}}&{\color{red}{23.1}}&{\color{red}{55.2}}&{\color{red}{81.0}}&{\color{red}{32.9}}&{\color{red}{31.4}}&{\color{red}{36.3}}&{\color{red}{28.6}}  \\ 
Next2Bit64&{\color{blue}{11.6}}&{\color{blue}{17.0}}&{\color{blue}{17.6}}&{\color{blue}{15.6}}&{\color{blue}{18.1}}&{\color{blue}{12.9}}&{\color{blue}{35.6}}&{\color{blue}{52.2}}&{\color{blue}{18.9}}&{\color{blue}{17.5}}&{\color{blue}{20.0}}&{\color{blue}{15.2}}  \\ 
\hline
NextBit32&{10.5} &{15.5} &{\color{red}{17.2}}&{\color{red}{15.2}}&{\color{red}{17.8}}&{\color{red}{12.6}}&{28.7} &{\color{blue}\emph{42.2}}&{\color{red}{19.4}}&{\color{red}{16.4}}&{\color{red}{30.4}}&{\color{red}{15.3}}  \\ 
NextBit32\_by\_mask&{10.3} &{15.2} &{\color{red}{17.6}}&{\color{red}{15.6}}&{\color{red}{18.2}}&{\color{red}{12.9}}&{28.7} &{\color{blue}\emph{42.4}}&{\color{red}{18.9}}&{\color{red}{16.8}}&{\color{red}{25.0}}&{\color{red}{15.5}}  \\ 
NextBit64&{\color{blue}{7.5}}&{\color{blue}{10.9}}&{\color{blue}{11.1}}&{\color{blue}{9.8}}&{\color{blue}{11.1}}&{\color{blue}{7.8}}&{\color{blue}{19.7}}&{\color{blue}{41.5}}&{\color{blue}{11.7}}&{\color{blue}{10.4}}&{\color{blue}{11.8}}&{\color{blue}{9.8}}  \\ 
\hline
NextByte32&{\color{red}{67.6}}&{\color{red}{99.5}}&{\color{red}{108.1}}&{\color{red}{95.7}}&{\color{red}{117.9}}&{\color{red}{84.8}}&{\color{red}{212.5}}&{\color{red}{315.5}}&{\color{red}{118.2}}&{\color{red}{113.7}}&{\color{red}{136.3}}&{\color{red}{102.7}}  \\ 
NextByte64&{\color{blue}\emph{36.2}}&{\color{blue}\emph{53.5}}&{\color{blue}{57.1}}&{\color{blue}{50.5}}&{\color{blue}{60.5}}&{\color{blue}{43.8}}&{\color{blue}{130.6}}&{\color{blue}{192.9}}&{\color{blue}{62.4}}&{\color{blue}\emph{59.7}}&{\color{blue}{69.2}}&{\color{blue}\emph{56.2}}  \\ 
NextByte64\_prefilled&{\color{blue}{35.7}}&{\color{blue}{52.5}}&{\color{blue}{56.9}}&{\color{blue}{50.3}}&{\color{blue}{60.6}}&{\color{blue}{43.7}}&{\color{blue}{131.0}}&{\color{blue}{192.7}}&{\color{blue}{62.6}}&{\color{blue}{56.4}}&{\color{blue}{69.7}}&{\color{blue}{54.1}}  \\ 
\hline
NextCard32&{56.2} &{83.3} &{\color{red}{88.1}}&{\color{red}{77.9}}&{\color{red}{96.5}}&{\color{red}{71.7}}&{173.9} &{255.6} &{\color{red}{97.9}}&{\color{red}{85.9}}&{\color{red}{110.4}}&{\color{red}{86.4}}  \\ 
NextCard32\_prefilled&{60.7} &{89.7} &{\color{red}{104.3}}&{\color{red}{92.2}}&{\color{red}{102.9}}&{\color{blue}\emph{37.6}}&{176.3} &{259.6} &{\color{red}{115.6}}&{\color{red}{106.3}}&{\color{red}{117.5}}&{45.3}   \\ 
NextCard64&{\color{blue}{46.4}}&{\color{blue}{68.1}}&{\color{blue}{43.5}}&{\color{blue}{38.4}}&{\color{blue}{49.1}}&{\color{blue}{34.8}}&{\color{blue}{119.0}}&{193.4} &{\color{blue}{48.5}}&{\color{blue}{43.1}}&{\color{blue}{55.3}}&{\color{blue}{40.7}}  \\ 
NextCard64\_prefilled&{\color{blue}{46.0}}&{\color{blue}{67.5}}&{\color{red}{74.7}}&{\color{red}{66.0}}&{62.0} &{\color{blue}\emph{35.2}}&{\color{blue}{118.7}}&{\color{blue}{173.6}}&{\color{red}{79.2}}&{\color{red}{72.9}}&{68.1} &{\color{blue}\emph{43.8}}  \\ 
\hline
NextWord32&{\color{red}{132.7}}&{\color{red}{195.7}}&{\color{red}{209.2}}&{\color{red}{185.4}}&{\color{red}{228.3}}&{\color{red}{166.7}}&{\color{red}{429.4}}&{\color{red}{633.0}}&{\color{red}{236.0}}&{\color{red}{204.2}}&{\color{red}{265.3}}&{\color{red}{202.9}}  \\ 
NextWord64&{\color{blue}{69.3}}&{\color{blue}{102.8}}&{\color{blue}{108.1}}&{\color{blue}{95.6}}&{\color{blue}{117.8}}&{\color{blue}{84.5}}&{\color{blue}{264.2}}&{\color{blue}{387.4}}&{\color{blue}{117.9}}&{\color{blue}{105.6}}&{\color{blue}{136.2}}&{\color{blue}{108.7}}  \\ 
\hline
my\_gen\_rand32&{\color{blue}{262.5}}&{\color{blue}{391.5}}&{\color{blue}{413.3}}&{\color{blue}{366.7}}&{\color{blue}{457.3}}&{\color{blue}{328.7}}&{\color{blue}{859.8}}&{\color{blue}{1263.3}}&{\color{blue}\emph{459.2}}&{\color{blue}{403.1}}&{\color{blue}{525.4}}&{\color{blue}{399.5}}  \\ 
my\_gen\_rand64&{\color{blue}{263.0}}&{\color{blue}{390.4}}&{\color{blue}{413.1}}&{\color{blue}{368.4}}&{\color{blue}{457.7}}&{\color{blue}{328.4}}&{1018.1} &{1513.1} &{\color{blue}{454.3}}&{\color{blue}{402.9}}&{\color{blue}{523.9}}&{\color{blue}{398.5}}  \\ 
\end{tabular}

\noindent\begin{tabular}{l|rrrrrr|rrrrrr|}
 & \multicolumn{6}{|c|}{counter}  \\
 &  HW1& HW2& HW3& HW4& HW5& HW6 \\
\hline
Next2Bit32&{\color{blue}{2.5}}&{\color{blue}{3.6}}&{\color{blue}\emph{4.6}}&{\color{blue}\emph{4.1}}&{\color{blue}\emph{3.6}}&{\color{blue}\emph{2.4}}&~ &~ &~ &~ &~ &~   \\ 
Next2Bit64&{3.4} &{4.9} &{\color{blue}{4.4}}&{\color{blue}{3.9}}&{\color{blue}{3.6}}&{\color{blue}{2.4}}&~ &~ &~ &~ &~ &~   \\ 
\hline
NextBit32&{\color{blue}{2.4}}&{\color{blue}{3.6}}&{\color{blue}\emph{4.3}}&{\color{blue}\emph{3.8}}&{\color{blue}{3.6}}&{\color{blue}{2.4}}&~ &~ &~ &~ &~ &~   \\ 
NextBit32\_by\_mask&{\color{blue}{2.4}}&{\color{blue}{3.6}}&{\color{blue}\emph{4.4}}&{\color{blue}\emph{3.9}}&{\color{blue}{3.6}}&{\color{blue}{2.4}}&~ &~ &~ &~ &~ &~   \\ 
NextBit64&{3.1} &{4.6} &{\color{blue}{4.2}}&{\color{blue}{3.7}}&{\color{blue}{3.6}}&{2.7} &~ &~ &~ &~ &~ &~   \\ 
\hline
NextByte32&{\color{blue}\emph{2.9}}&{4.3} &{\color{blue}{4.8}}&{\color{blue}{4.2}}&{4.3} &{\color{blue}{2.8}}&~ &~ &~ &~ &~ &~   \\ 
NextByte64&{\color{blue}\emph{2.9}}&{4.3} &{5.3} &{4.7} &{\color{blue}\emph{3.8}}&{\color{blue}\emph{2.9}}&~ &~ &~ &~ &~ &~   \\ 
NextByte64\_prefilled&{\color{blue}{2.8}}&{\color{blue}{3.8}}&{\color{blue}\emph{5.2}}&{\color{blue}\emph{4.6}}&{\color{blue}{3.6}}&{\color{blue}\emph{3.0}}&~ &~ &~ &~ &~ &~   \\ 
\hline
NextCard32&{\color{blue}{2.8}}&{\color{blue}{4.2}}&{\color{blue}{5.1}}&{\color{blue}{4.5}}&{\color{blue}{4.5}}&{\color{red}{6.4}}&~ &~ &~ &~ &~ &~   \\ 
NextCard32\_prefilled&{\color{red}{5.2}}&{\color{red}{7.7}}&{\color{red}{21.9}}&{\color{red}{19.4}}&{\color{red}{8.9}}&{\color{red}{6.9}}&~ &~ &~ &~ &~ &~   \\ 
NextCard64&{\color{red}{22.0}}&{\color{red}{32.7}}&{6.0} &{5.4} &{6.0} &{\color{blue}{4.2}}&~ &~ &~ &~ &~ &~   \\ 
NextCard64\_prefilled&{\color{red}{21.7}}&{\color{red}{31.9}}&{\color{red}{36.9}}&{\color{red}{32.7}}&{\color{red}{19.7}}&{4.9} &~ &~ &~ &~ &~ &~   \\ 
\hline
NextWord32&{\color{blue}{2.9}}&{\color{blue}{4.3}}&{\color{blue}\emph{5.0}}&{\color{blue}\emph{4.4}}&{4.8} &{3.2} &~ &~ &~ &~ &~ &~   \\ 
NextWord64&{3.3} &{4.9} &{\color{blue}{4.6}}&{\color{blue}{4.1}}&{\color{blue}{4.0}}&{\color{blue}{2.7}}&~ &~ &~ &~ &~ &~   \\ 
\hline
my\_gen\_rand32&{\color{blue}{2.0}}&{\color{blue}{3.0}}&{\color{blue}{3.6}}&{\color{blue}{3.2}}&{\color{blue}{3.5}}&{\color{blue}{2.0}}&~ &~ &~ &~ &~ &~   \\ 
my\_gen\_rand64&{2.4} &{3.5} &{\color{blue}{3.6}}&{\color{blue}{3.2}}&{\color{blue}{3.5}}&{2.3} &~ &~ &~ &~ &~ &~   \\ 
\end{tabular}

}

\subsection{Graphs}
In all of the following graphs, the abscissa is $n$, (that is the
modulus of the URFs); the abscissa is in log-scale (precisely,
it contains all $n$ from 2 to 32, and then $n$ is incremented by
$\lfloor n/32\rfloor$ up to $2^{32}$, for a total of 733 samples). The
number in parentheses near the graph labels are the average time for
call (in nanoseconds; averaged in the aforementioned log scale).

\subsubsection{Uniform random functions}
\label{sec:graph_speeds}
To reduce the size of the labels, we abbreviated
\texttt{uniform\_random\_by\_bit\_recycling} as \texttt{bbr}, and
\texttt{uniform\_random\_simple} as \texttt{simple}.

\subsubsection*{SFMT}
\noindent
\includegraphics[width=0.45\linewidth]{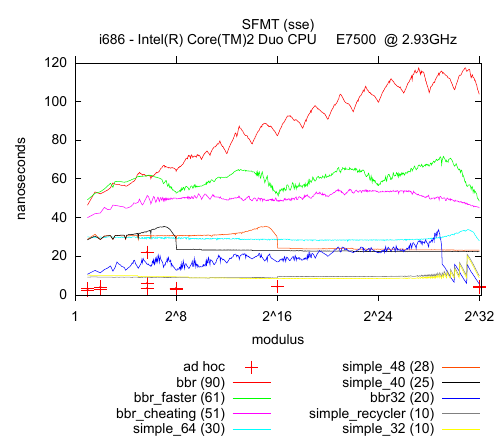}
\includegraphics[width=0.45\linewidth]{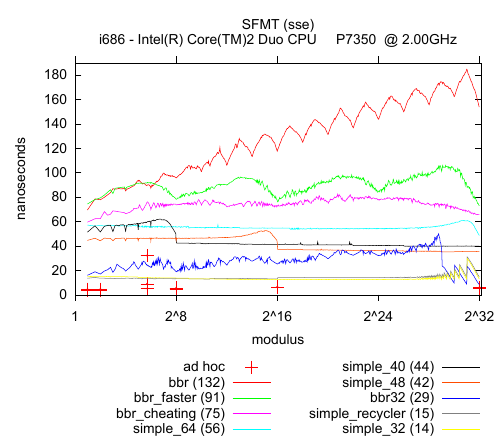}
\\\noindent
\includegraphics[width=0.45\linewidth]{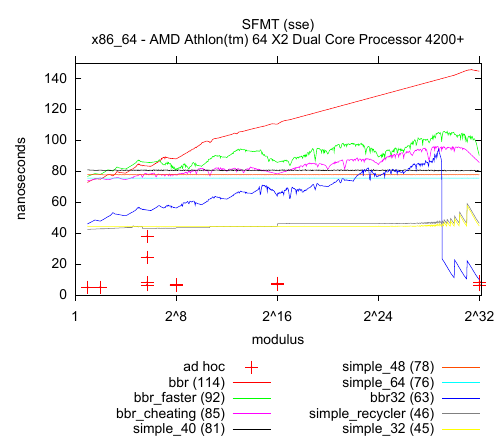}
\includegraphics[width=0.45\linewidth]{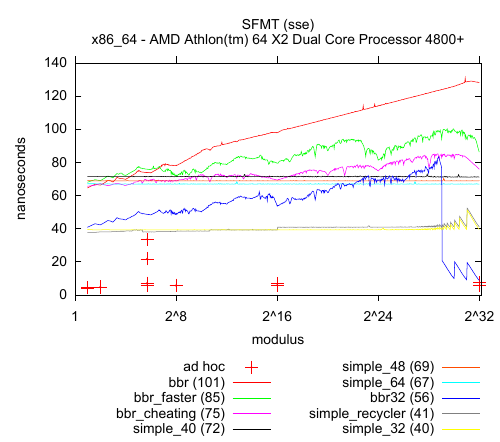}
\\\noindent
\includegraphics[width=0.45\linewidth]{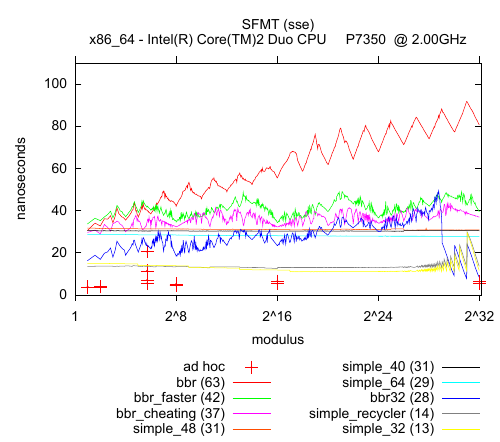}
\includegraphics[width=0.45\linewidth]{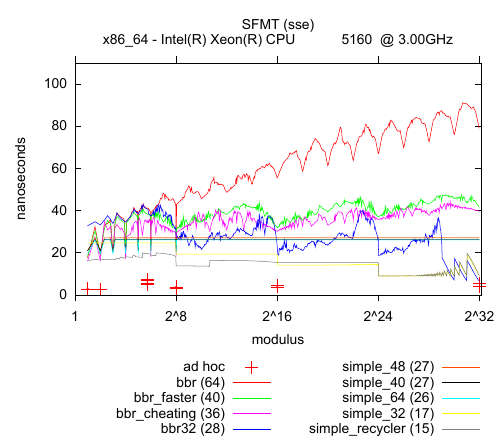}

\subsubsection*{xorshift}

\noindent
\includegraphics[width=0.45\linewidth]{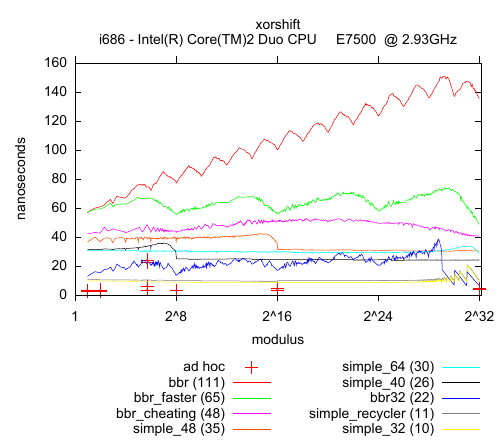}
\includegraphics[width=0.45\linewidth]{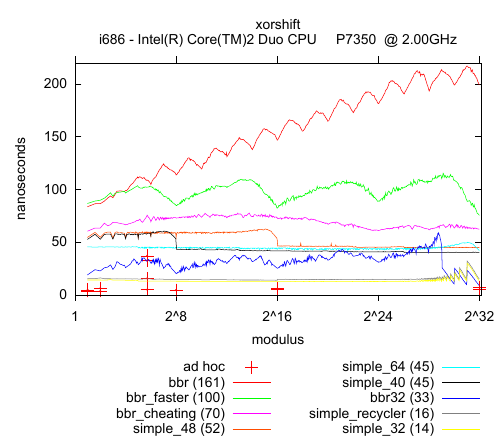}
\\\noindent
\includegraphics[width=0.45\linewidth]{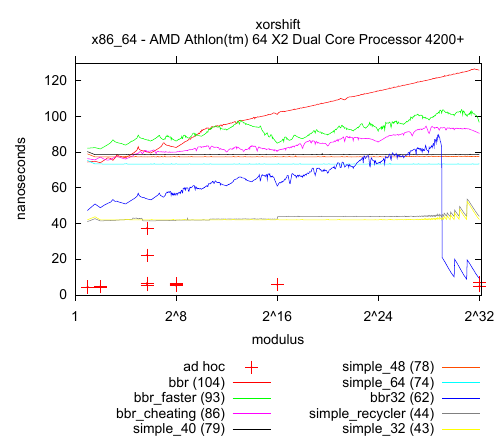}
\includegraphics[width=0.45\linewidth]{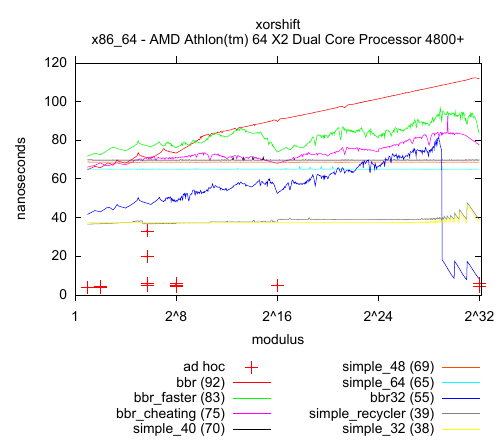}
\\\noindent
\includegraphics[width=0.45\linewidth]{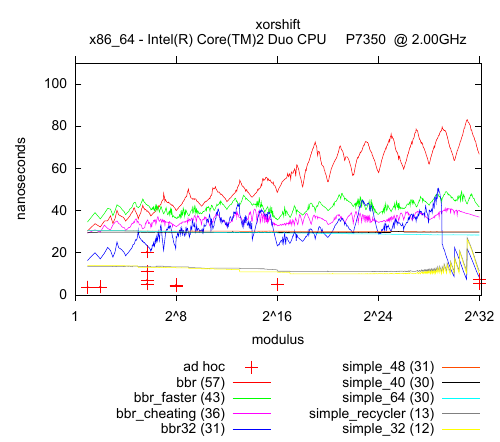}
\includegraphics[width=0.45\linewidth]{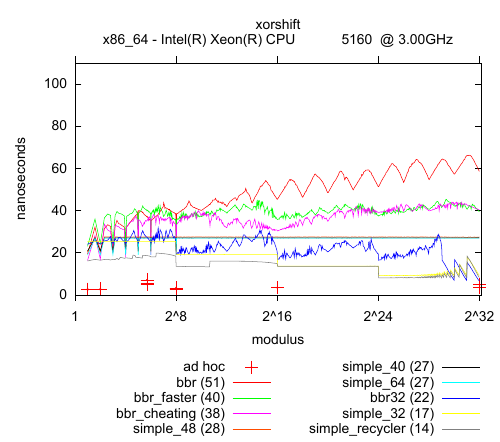}

\subsubsection*{SFMT + MD5}

\noindent
\includegraphics[width=0.45\linewidth]{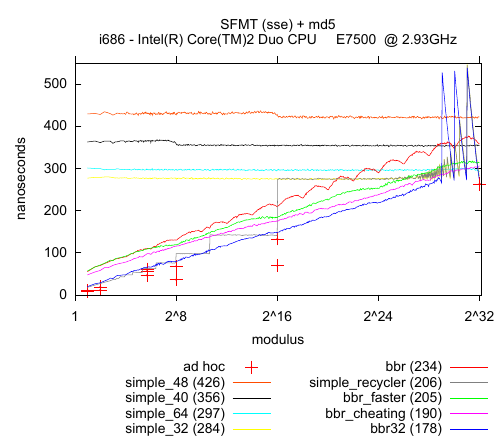}
\includegraphics[width=0.45\linewidth]{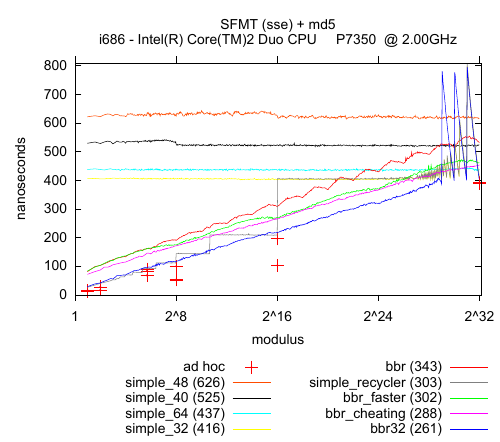}
\\\noindent
\includegraphics[width=0.45\linewidth]{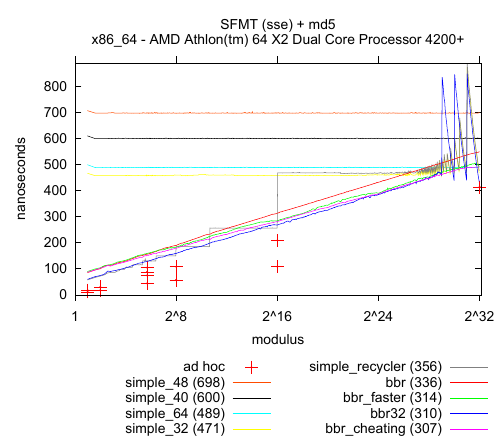}
\includegraphics[width=0.45\linewidth]{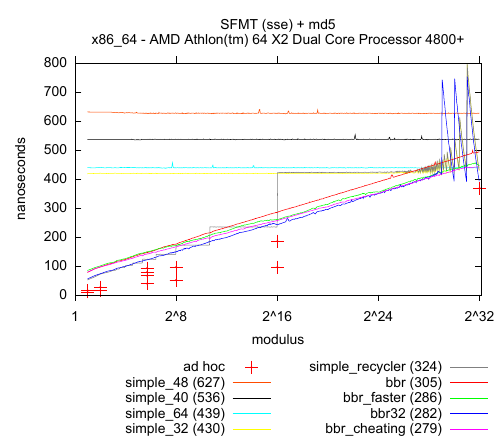}
\\\noindent
\includegraphics[width=0.45\linewidth]{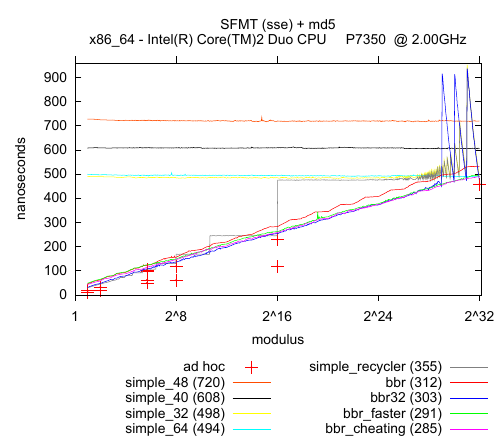}
\includegraphics[width=0.45\linewidth]{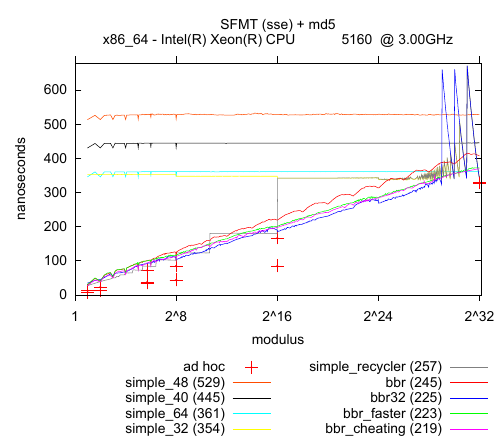}

\subsubsection*{bbs260}
\noindent
\includegraphics[width=0.45\linewidth]{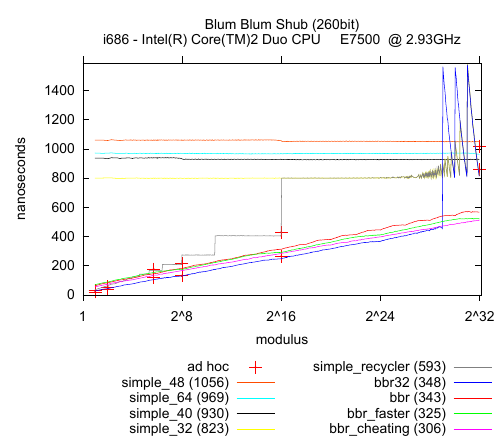}
\includegraphics[width=0.45\linewidth]{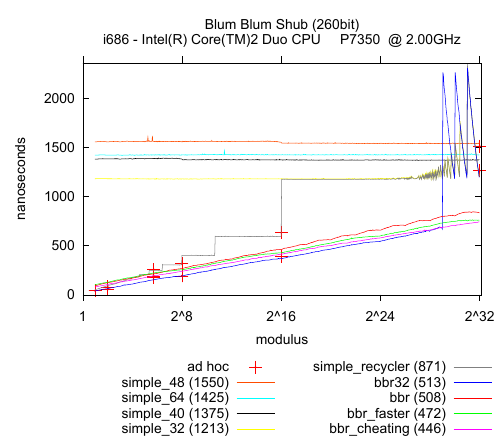}
\\\noindent
\includegraphics[width=0.45\linewidth]{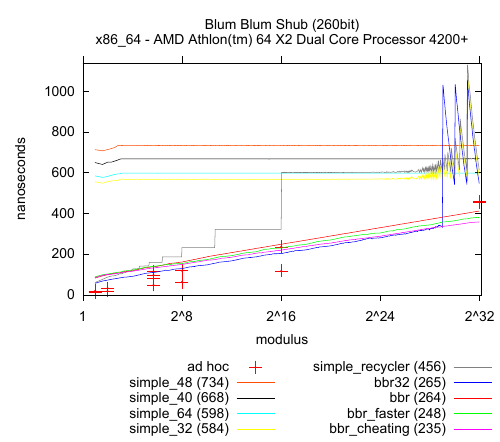}
\includegraphics[width=0.45\linewidth]{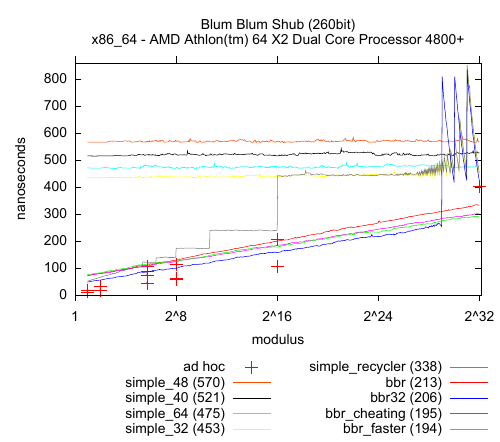}
\\\noindent
\includegraphics[width=0.45\linewidth]{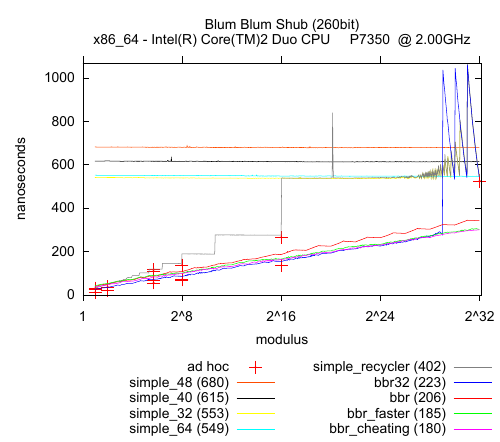}
\includegraphics[width=0.45\linewidth]{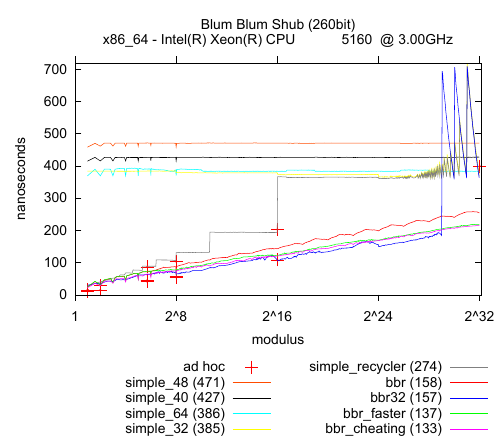}

\subsubsection{Integer arithmetic}
\label{sec:graph_arith_speeds}
\noindent\begin{minipage}[b]{0.49\linewidth}{\scriptsize%
\noindent%
\begin{verbatim}
typedef uint64_t st;
st div32(uint32_t n) {
  return my_gen_rand32() / n ;
}
st div32_24(uint32_t n) {
  return (my_gen_rand32() & 0xFF000000) / n ;
}
st div48(uint32_t n) {
  uint64_t r = my_gen_rand32(), m = r << 16;
  return m / n ;
}
st div64(uint32_t n) {
  uint64_t m = my_gen_rand64() / n ;
  return m;
}
st mod32(uint32_t n) {
  return my_gen_rand32() % n ;
}
st mod32_24(uint32_t n) {
  return (my_gen_rand32() & 0xFF000000) % n ;
}
st mod48(uint32_t n) {
  uint64_t r = my_gen_rand32(), m = r << 16;
  return m % n ;
}
st mod64(uint32_t n) {
  uint64_t m =my_gen_rand64() % n ;
  return m;
}
st prod32(uint32_t n) {
  return my_gen_rand32() * n ;
}
st prod32_24(uint32_t n) {
  return (my_gen_rand32() & 0xFF) * n ;
}
st prod48(uint32_t n) {
  uint64_t r = my_gen_rand32(), m = r << 16;
  return m * n ;
}
st prod64(uint32_t n) {
  uint64_t m = my_gen_rand64() * n ;
  return m;
}
\end{verbatim}
}\noindent%
\includegraphics[width=0.99\linewidth]{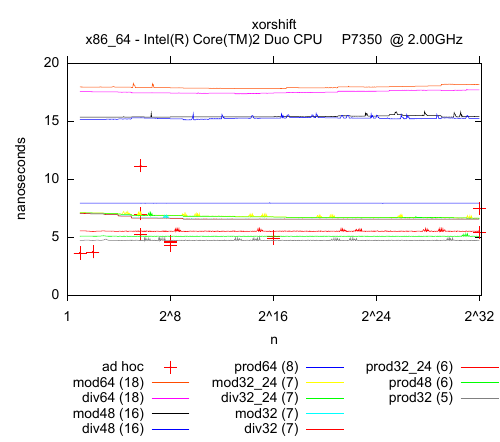}
\end{minipage}
\begin{minipage}[b]{0.48\linewidth}
\noindent
\includegraphics[width=0.99\linewidth]{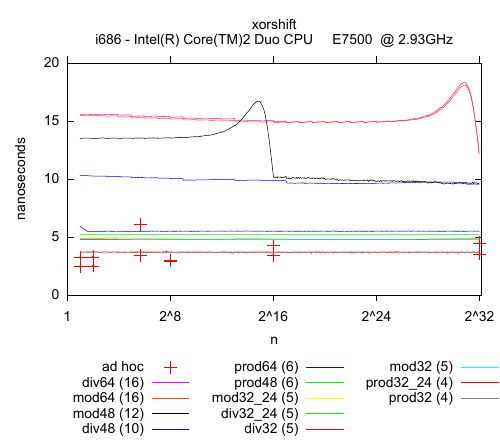}
\\\noindent
\includegraphics[width=0.99\linewidth]{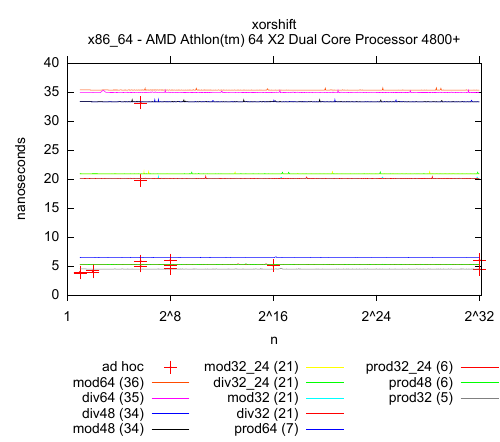}
\\\noindent
\includegraphics[width=0.99\linewidth]{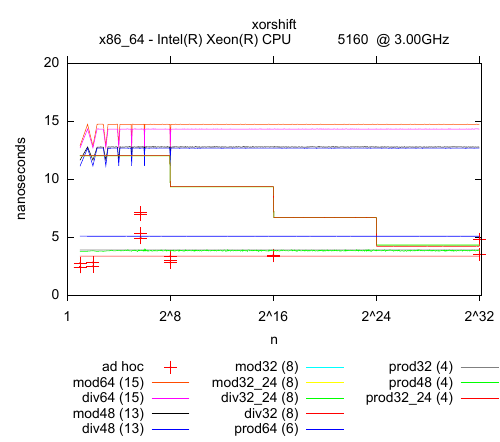}
\end{minipage}

\section[C code]{\proglang{C} Code}
\label{sec:C_code}
The code that we wrote
 is licensed according to the \emph{Gnu Public License v2.0}.


\emph{(In the following, we describe only the code in the three most
significant files. All the code is documented with
Doxygen, and the complete code documentation
is available in \texttt{doc/refman.pdf} or
\texttt{doc/html/index.html} in the source code tarball.
Those documents also explain how to compile the code and
run the numerical tests.)}

\subsection{Back-end RNGs}
\label{sec:backend_C}


\def\hyperlink#1{}

{\small
\hypertarget{_r_n_g_8c}{
\subsubsection{RNG.c File Reference}
\label{_r_n_g_8c}\index{RNG.c@{RNG.c}}
}

Backend RNG wrapper.

{\ttfamily \#include \char`\"{}SFMT.h\char`\"{}}\par
\subsubsection*{Defines}
\begin{DoxyCompactItemize}
\item 
\#define \hyperlink{_r_n_g_8c_aec13afa2a978e714e40279ce5538de07}{COUNTBITS}(A)
\item 
\#define \hyperlink{_r_n_g_8c_a5b0885b8b55bbc13691092b704d9309f}{RNG}~1
\end{DoxyCompactItemize}
\subsubsection*{Functions}
\begin{DoxyCompactItemize}
\item 
uint32\textunderscore{}t \hyperlink{_r_n_g_8c_affb62bf72df7716e14bee3cdda966283}{my\textunderscore{}gen\textunderscore{}rand32} ()
\item 
uint64\textunderscore{}t \hyperlink{_r_n_g_8c_affcfff41ea70b785738c2911275830b3}{my\textunderscore{}gen\textunderscore{}rand64} ()
\item 
void \hyperlink{_r_n_g_8c_aa300743dfcb26c0197761f867497e966}{my\textunderscore{}init\textunderscore{}gen\textunderscore{}rand} (uint32\textunderscore{}t seed)
\end{DoxyCompactItemize}
\subsubsection*{Variables}
\begin{DoxyCompactItemize}
\item 
\hypertarget{_r_n_g_8c_a8d642363434e7b276aaf7eccc30911e4}{
char $\ast$ \hyperlink{_r_n_g_8c_a8d642363434e7b276aaf7eccc30911e4}{RNGNAME} = \char`\"{}SFMT\char`\"{}}
\label{_r_n_g_8c_a8d642363434e7b276aaf7eccc30911e4}

\begin{DoxyCompactList}\small\item\em RNG name , as C string. \end{DoxyCompactList}\item 
\hypertarget{_r_n_g_8c_a15e458279deb160459720a6813c0de83}{
char $\ast$ \hyperlink{_r_n_g_8c_a15e458279deb160459720a6813c0de83}{RNGNICK} = \char`\"{}sfmt\char`\"{}}
\label{_r_n_g_8c_a15e458279deb160459720a6813c0de83}

\begin{DoxyCompactList}\small\item\em RNG nickname , as C string (used inside filenames) \end{DoxyCompactList}\end{DoxyCompactItemize}

\subsubsection{Detailed Description}
This code wraps up the backend RNG, and provides standardized calls that return a 32bit or 64bit random number.

To compile this code, define the RNG and COUNTBITS preprocessor macros, as explained below. 

\subsubsection{Define Documentation}
\hypertarget{_r_n_g_8c_aec13afa2a978e714e40279ce5538de07}{
\index{RNG.c@{RNG.c}!COUNTBITS@{COUNTBITS}}
\index{COUNTBITS@{COUNTBITS}!RNG.c@{RNG.c}}
\DoxySubstituteParagraph[{COUNTBITS}]{\setlength{\rightskip}{0pt plus 5cm}\#define COUNTBITS(
\begin{DoxyParamCaption}
\item[{}]{A}
\end{DoxyParamCaption}
)}}
\label{_r_n_g_8c_aec13afa2a978e714e40279ce5538de07}
is defined either as

\begin{DoxyItemize}
\item \#define \hyperlink{_r_n_g_8c_aec13afa2a978e714e40279ce5538de07}{COUNTBITS(A)}
\end{DoxyItemize}

to disable bit counting, or as something like

\begin{DoxyItemize}
\item \#define \hyperlink{_r_n_g_8c_aec13afa2a978e714e40279ce5538de07}{COUNTBITS(A)} input\textunderscore{}bits+=(A);
\end{DoxyItemize}

where input\textunderscore{}bits is a global (unsigned long) variable. \hypertarget{_r_n_g_8c_a5b0885b8b55bbc13691092b704d9309f}{
\index{RNG.c@{RNG.c}!RNG@{RNG}}
\index{RNG@{RNG}!RNG.c@{RNG.c}}
\DoxySubstituteParagraph[{RNG}]{\setlength{\rightskip}{0pt plus 5cm}\#define RNG~1}}
\label{_r_n_g_8c_a5b0885b8b55bbc13691092b704d9309f}
Choose the RNG backend by setting -\/DRNG=n where n is:

\begin{DoxyItemize}
\item 1 -\/$>$ SFMT , by M. Saito and M. Matsumoto;
\end{DoxyItemize}

\begin{DoxyItemize}
\item 2 -\/$>$ SFMT + md5 ;
\end{DoxyItemize}

\begin{DoxyItemize}
\item 3 -\/$>$ xorshift , by Marsaglia ;
\end{DoxyItemize}

\begin{DoxyItemize}
\item 4 -\/$>$ Blum Blum Shub with $\sim$128bit (product of two $\sim$31bit primes) (only on amd64, using gcc 128 int types) ;
\end{DoxyItemize}

\begin{DoxyItemize}
\item 5 -\/$>$ Blum Blum Shub with $\sim$260bit modulus (product of two $\sim$130bit primes);
\end{DoxyItemize}

(disclaimer: methods 2,4,5 are not guaranteed to generate high quality pseudonumbers; they were used only to test the code speed).

(Note that the documentation is generated assuming that RNG=1) 

\subsubsection{Function Documentation}
\hypertarget{_r_n_g_8c_affb62bf72df7716e14bee3cdda966283}{
\index{RNG.c@{RNG.c}!my gen rand32@{my gen rand32}}
\index{my gen rand32@{my gen rand32}!RNG.c@{RNG.c}}
\DoxySubstituteParagraph[{my\textunderscore{}gen\textunderscore{}rand32}]{\setlength{\rightskip}{0pt plus 5cm}uint32\textunderscore{}t my\textunderscore{}gen\textunderscore{}rand32 (
\begin{DoxyParamCaption}
{}
\end{DoxyParamCaption}
)}}
\label{_r_n_g_8c_affb62bf72df7716e14bee3cdda966283}
standardized call that returns a 32bit random number 
\begin{DoxyCode}
\{
  uint32\_t r=gen\_rand32();
  \hyperlink{_r_n_g_8c_aec13afa2a978e714e40279ce5538de07}{COUNTBITS}(32);
  \textcolor{keywordflow}{return} r;
\}
\end{DoxyCode}
\hypertarget{_r_n_g_8c_affcfff41ea70b785738c2911275830b3}{
\index{RNG.c@{RNG.c}!my gen rand64@{my gen rand64}}
\index{my gen rand64@{my gen rand64}!RNG.c@{RNG.c}}
\DoxySubstituteParagraph[{my\textunderscore{}gen\textunderscore{}rand64}]{\setlength{\rightskip}{0pt plus 5cm}uint64\textunderscore{}t my\textunderscore{}gen\textunderscore{}rand64 (
\begin{DoxyParamCaption}
{}
\end{DoxyParamCaption}
)}}
\label{_r_n_g_8c_affcfff41ea70b785738c2911275830b3}
standardized call that returns a 64bit random number 
\begin{DoxyCode}
\{
  uint64\_t r=gen\_rand64();
  \hyperlink{_r_n_g_8c_aec13afa2a978e714e40279ce5538de07}{COUNTBITS}(64);
  \textcolor{keywordflow}{return} r;
\}
\end{DoxyCode}
\hypertarget{_r_n_g_8c_aa300743dfcb26c0197761f867497e966}{
\index{RNG.c@{RNG.c}!my init gen rand@{my init gen rand}}
\index{my init gen rand@{my init gen rand}!RNG.c@{RNG.c}}
\DoxySubstituteParagraph[{my\textunderscore{}init\textunderscore{}gen\textunderscore{}rand}]{\setlength{\rightskip}{0pt plus 5cm}void my\textunderscore{}init\textunderscore{}gen\textunderscore{}rand (
\begin{DoxyParamCaption}
\item[{uint32\textunderscore{}t}]{seed}
\end{DoxyParamCaption}
)}}
\label{_r_n_g_8c_aa300743dfcb26c0197761f867497e966}
standardized call that sets the random seed 
\begin{DoxyCode}
\{
  init\_gen\_rand(seed);
\}
\end{DoxyCode}

}



\subsection{{ad hoc} functions}
\label{sec:adhoc_C}

{\small
  \hypertarget{AH8c}{
\subsubsection{ad hoc.c File Reference}
\label{AH8c}\index{ad hoc.c@{ad hoc.c}}
}

Ad-\/hoc functions that return some random bits.

\subsubsection*{Functions}
\begin{DoxyCompactItemize}
\item 
unsigned int \hyperlink{AH8c_a437e1e0a6509b6dd858d1be826315993}{NextByte32} ()
\item 
unsigned int \hyperlink{AH8c_a10ccbd8b2b9af41fce4a5d40640c90a7}{NextByte64} ()
\item 
unsigned int \hyperlink{AH8c_a5af345aa5975099d29e7847c20343f40}{NextByte64\textunderscore{}prefilled} ()
\item 
unsigned int \hyperlink{AH8c_a1d45cfcf6adada5e5c78463dd703a4b7}{NextWord32} ()
\item 
unsigned int \hyperlink{AH8c_a9c0eee176aba60fd9c925fed0b93722a}{NextWord64} ()
\item 
unsigned int \hyperlink{AH8c_a8851d9e47b8825a0629a98adbbea1035}{Next2Bit32} ()
\item 
unsigned int \hyperlink{AH8c_afac0a831fa0c744f99e977f585c6017f}{Next2Bit64} ()
\item 
unsigned int \hyperlink{AH8c_a428968c57aa235b0e5119cacb9bdf378}{NextBit32\textunderscore{}by\textunderscore{}mask} ()
\item 
unsigned int \hyperlink{AH8c_a25b4b4c818d81e87adeefc66a576667c}{NextBit32} ()
\item 
unsigned int \hyperlink{AH8c_a96bd20b43ef56a94ac5574b416d5e6e7}{NextBit64} ()
\item 
unsigned int \hyperlink{AH8c_ad59574b55d47cec0ad8f9b18f2172608}{NextCard32} ()
\item 
unsigned int \hyperlink{AH8c_a64999255eeed624ae795239adc9cda90}{NextCard64} ()
\item 
unsigned int \hyperlink{AH8c_a740a5a58d1613efb788b8d91c17845cb}{NextCard32\textunderscore{}prefilled} ()
\item 
unsigned int \hyperlink{AH8c_aca6c7a9cd64b3924bcc52b2d1ff85ebc}{NextCard64\textunderscore{}prefilled} ()
\end{DoxyCompactItemize}
\subsubsection*{Variables}
\begin{DoxyCompactItemize}
\item 
\hypertarget{AH8c_a9b7d4e84836e8d0c811ea05520b60572}{
unsigned int {\bfseries \textunderscore{}\textunderscore{}saved\textunderscore{}bytes} \mbox{[}8\mbox{]}}
\label{AH8c_a9b7d4e84836e8d0c811ea05520b60572}

\item 
\hypertarget{AH8c_a91c4b1c606df3a41b50f1d837456b227}{
unsigned char {\bfseries \textunderscore{}\textunderscore{}32saved\textunderscore{}cards} \mbox{[}5\mbox{]}}
\label{AH8c_a91c4b1c606df3a41b50f1d837456b227}

\item 
\hypertarget{AH8c_a3332dbad5ceaf937c628432db20b788e}{
unsigned char {\bfseries \textunderscore{}\textunderscore{}64saved\textunderscore{}cards} \mbox{[}11\mbox{]}}
\label{AH8c_a3332dbad5ceaf937c628432db20b788e}

\end{DoxyCompactItemize}

\subsubsection{Detailed Description}
Contains the ad\textunderscore{}hoc functions; each ad\textunderscore{}hoc function returns some random bits; there are many different implementations of each function; then the 'optimize' function will benchmark them and prepare a specific header that selects the fastest one (given the current architecture and the chosen RNG). See \hyperlink{optimizer__sfmt__sse__fixed_8h}{optimizer\textunderscore{}sfmt\textunderscore{}sse\textunderscore{}fixed.h} for an example such header. 

\subsubsection{Function Documentation}
\hypertarget{AH8c_a8851d9e47b8825a0629a98adbbea1035}{
\index{ad hoc.c@{ad hoc.c}!Next2Bit32@{Next2Bit32}}
\index{Next2Bit32@{Next2Bit32}!ad_hoc.c@{ad hoc.c}}
\DoxySubstituteParagraph[{Next2Bit32}]{\setlength{\rightskip}{0pt plus 5cm}unsigned int Next2Bit32 (
\begin{DoxyParamCaption}
{}
\end{DoxyParamCaption}
)}}
\label{AH8c_a8851d9e47b8825a0629a98adbbea1035}
returns a random number in the range 0...3 , using a 32bit generator. 
\begin{DoxyCode}
\{
  \textcolor{keyword}{static} \textcolor{keywordtype}{int} l=0;
  \textcolor{keyword}{static} uint32\_t R=0;
  \textcolor{keywordflow}{if}(\hyperlink{utils_8h_ac6c45889010c1bd68631771b64f18101}{unlikely}(l<=0))\{
    R=\hyperlink{_r_n_g_8c_affb62bf72df7716e14bee3cdda966283}{my_gen_rand32}();
    l=16;
  \}
  \textcolor{keywordtype}{unsigned} \textcolor{keywordtype}{int} bit=R&3;
  R>>=2;
  l--;
  \textcolor{keywordflow}{return} bit;
\}
\end{DoxyCode}
\hypertarget{AH8c_afac0a831fa0c744f99e977f585c6017f}{
\index{ad hoc.c@{ad hoc.c}!Next2Bit64@{Next2Bit64}}
\index{Next2Bit64@{Next2Bit64}!ad_hoc.c@{ad hoc.c}}
\DoxySubstituteParagraph[{Next2Bit64}]{\setlength{\rightskip}{0pt plus 5cm}unsigned int Next2Bit64 (
\begin{DoxyParamCaption}
{}
\end{DoxyParamCaption}
)}}
\label{AH8c_afac0a831fa0c744f99e977f585c6017f}
returns a random number in the range 0...3 , using a 64bit generator. 
\begin{DoxyCode}
\{
  \textcolor{keyword}{static} \textcolor{keywordtype}{int} l=0;
  \textcolor{keyword}{static} uint64\_t R=0;
  \textcolor{keywordflow}{if}(\hyperlink{utils_8h_ac6c45889010c1bd68631771b64f18101}{unlikely}(l<=0))\{
    R=\hyperlink{_r_n_g_8c_affcfff41ea70b785738c2911275830b3}{my_gen_rand64}();
    l=32;
  \}
  \textcolor{keywordtype}{unsigned} \textcolor{keywordtype}{int} bit=R&3;
  R>>=2;
  l--;
  \textcolor{keywordflow}{return} bit;
\}
\end{DoxyCode}
\hypertarget{AH8c_a25b4b4c818d81e87adeefc66a576667c}{
\index{ad hoc.c@{ad hoc.c}!NextBit32@{NextBit32}}
\index{NextBit32@{NextBit32}!ad_hoc.c@{ad hoc.c}}
\DoxySubstituteParagraph[{NextBit32}]{\setlength{\rightskip}{0pt plus 5cm}unsigned int NextBit32 (
\begin{DoxyParamCaption}
{}
\end{DoxyParamCaption}
)}}
\label{AH8c_a25b4b4c818d81e87adeefc66a576667c}
returns a random bit, using a 32bit generator. 
\begin{DoxyCode}
\{
  \textcolor{keyword}{static} \textcolor{keywordtype}{int} l=0;
  \textcolor{keyword}{static} uint32\_t R=0;
  \textcolor{keywordflow}{if}(\hyperlink{utils_8h_ac6c45889010c1bd68631771b64f18101}{unlikely}(l<=0))\{
    R=\hyperlink{_r_n_g_8c_affb62bf72df7716e14bee3cdda966283}{my_gen_rand32}();
    l=32;
  \}
  \textcolor{keywordtype}{unsigned} \textcolor{keywordtype}{int} bit=R&1;
  R>>=1;
  l--;
  \textcolor{keywordflow}{return} bit;
\}
\end{DoxyCode}
\hypertarget{AH8c_a428968c57aa235b0e5119cacb9bdf378}{
\index{ad hoc.c@{ad hoc.c}!NextBit32 by mask@{NextBit32 by mask}}
\index{NextBit32 by mask@{NextBit32 by mask}!ad_hoc.c@{ad hoc.c}}
\DoxySubstituteParagraph[{NextBit32\textunderscore{}by\textunderscore{}mask}]{\setlength{\rightskip}{0pt plus 5cm}unsigned int NextBit32\textunderscore{}by\textunderscore{}mask (
\begin{DoxyParamCaption}
{}
\end{DoxyParamCaption}
)}}
\label{AH8c_a428968c57aa235b0e5119cacb9bdf378}
returns a random bit, using a 32bit generator. 
\begin{DoxyCode}
\{
  \textcolor{keyword}{static} \textcolor{keywordtype}{int} l=0;
  \textcolor{keyword}{static} uint32\_t R=0;
  \textcolor{keywordflow}{if}(\hyperlink{utils_8h_ac6c45889010c1bd68631771b64f18101}{unlikely}(l<=0))\{
    R=\hyperlink{_r_n_g_8c_affb62bf72df7716e14bee3cdda966283}{my_gen_rand32}();
    l=32;
  \}
  l--;
  \textcolor{keywordflow}{return} (R & bitmasks[l]) ? 1 : 0;
\}
\end{DoxyCode}
\hypertarget{AH8c_a96bd20b43ef56a94ac5574b416d5e6e7}{
\index{ad hoc.c@{ad hoc.c}!NextBit64@{NextBit64}}
\index{NextBit64@{NextBit64}!ad_hoc.c@{ad hoc.c}}
\DoxySubstituteParagraph[{NextBit64}]{\setlength{\rightskip}{0pt plus 5cm}unsigned int NextBit64 (
\begin{DoxyParamCaption}
{}
\end{DoxyParamCaption}
)}}
\label{AH8c_a96bd20b43ef56a94ac5574b416d5e6e7}
returns a random bit, using a 64bit generator. 
\begin{DoxyCode}
\{
  \textcolor{keyword}{static} \textcolor{keywordtype}{int} l=0;
  \textcolor{keyword}{static} uint64\_t R=0;
  \textcolor{keywordflow}{if}(\hyperlink{utils_8h_ac6c45889010c1bd68631771b64f18101}{unlikely}(l<=0))\{
    R=\hyperlink{_r_n_g_8c_affcfff41ea70b785738c2911275830b3}{my_gen_rand64}();
    l=64;
  \}
  \textcolor{keywordtype}{unsigned} \textcolor{keywordtype}{int} bit=R&1;
  R>>=1;
  l--;
  \textcolor{keywordflow}{return} bit;
\}
\end{DoxyCode}
\hypertarget{AH8c_a437e1e0a6509b6dd858d1be826315993}{
\index{ad hoc.c@{ad hoc.c}!NextByte32@{NextByte32}}
\index{NextByte32@{NextByte32}!ad_hoc.c@{ad hoc.c}}
\DoxySubstituteParagraph[{NextByte32}]{\setlength{\rightskip}{0pt plus 5cm}unsigned int NextByte32 (
\begin{DoxyParamCaption}
{}
\end{DoxyParamCaption}
)}}
\label{AH8c_a437e1e0a6509b6dd858d1be826315993}
returns a random number in the range 0...255, using a 32bit generator. 
\begin{DoxyCode}
\{
  \textcolor{keyword}{static} \textcolor{keywordtype}{int} l=0;
  \textcolor{keyword}{static} uint32\_t R=0;
  \textcolor{keywordflow}{if}(\hyperlink{utils_8h_ac6c45889010c1bd68631771b64f18101}{unlikely}(l<=0))\{
    R=\hyperlink{_r_n_g_8c_affb62bf72df7716e14bee3cdda966283}{my_gen_rand32}();
    l=4;
  \}
  \textcolor{keywordtype}{unsigned} \textcolor{keywordtype}{int} byte=R&255;
  l--;
  \textcolor{keywordflow}{if}(l) R>>=8;
  \textcolor{keywordflow}{return} byte;
\}
\end{DoxyCode}
\hypertarget{AH8c_a10ccbd8b2b9af41fce4a5d40640c90a7}{
\index{ad hoc.c@{ad hoc.c}!NextByte64@{NextByte64}}
\index{NextByte64@{NextByte64}!ad_hoc.c@{ad hoc.c}}
\DoxySubstituteParagraph[{NextByte64}]{\setlength{\rightskip}{0pt plus 5cm}unsigned int NextByte64 (
\begin{DoxyParamCaption}
{}
\end{DoxyParamCaption}
)}}
\label{AH8c_a10ccbd8b2b9af41fce4a5d40640c90a7}
returns a random number in the range 0...255, using a 64bit generator. 
\begin{DoxyCode}
\{
  \textcolor{keyword}{static} \textcolor{keywordtype}{int} l=0;
  \textcolor{keyword}{static} uint64\_t R=0;
  \textcolor{keywordflow}{if}(\hyperlink{utils_8h_ac6c45889010c1bd68631771b64f18101}{unlikely}(l<=0))\{
    R=\hyperlink{_r_n_g_8c_affcfff41ea70b785738c2911275830b3}{my_gen_rand64}();
    l=8;
  \}
  \textcolor{keywordtype}{unsigned} \textcolor{keywordtype}{int} byte=R & 255;
  l--;
  \textcolor{keywordflow}{if}(l) R>>=8;
  \textcolor{keywordflow}{return} byte;
\}
\end{DoxyCode}
\hypertarget{AH8c_a5af345aa5975099d29e7847c20343f40}{
\index{ad hoc.c@{ad hoc.c}!NextByte64 prefilled@{NextByte64 prefilled}}
\index{NextByte64 prefilled@{NextByte64 prefilled}!ad_hoc.c@{ad hoc.c}}
\DoxySubstituteParagraph[{NextByte64\textunderscore{}prefilled}]{\setlength{\rightskip}{0pt plus 5cm}unsigned int NextByte64\textunderscore{}prefilled (
\begin{DoxyParamCaption}
{}
\end{DoxyParamCaption}
)}}
\label{AH8c_a5af345aa5975099d29e7847c20343f40}
returns a random number in the range 0-\/255, using a 32bit generator, and an internal state. 
\begin{DoxyCode}
\{
  \textcolor{keyword}{static} \textcolor{keywordtype}{int} l=0;
  \textcolor{keywordflow}{if}(\hyperlink{utils_8h_ac6c45889010c1bd68631771b64f18101}{unlikely}(l<=0))\{
    uint64\_t R=\hyperlink{_r_n_g_8c_affcfff41ea70b785738c2911275830b3}{my_gen_rand64}();
    \_\_saved\_bytes[0] = R & 255;
    R >>= 8;
    \_\_saved\_bytes[1] = R & 255;
    R >>= 8;
    \_\_saved\_bytes[2] = R & 255;
    R >>= 8;
    \_\_saved\_bytes[3] = R & 255;
    R >>= 8;
    \_\_saved\_bytes[4] = R & 255;
    R >>= 8;
    \_\_saved\_bytes[5] = R & 255;
    R >>= 8;
    \_\_saved\_bytes[6] = R & 255;
    R >>= 8;
    \_\_saved\_bytes[7] = R & 255;
    l=8;
  \}
  l--;
  \textcolor{keywordflow}{return} \_\_saved\_bytes[l];
\}
\end{DoxyCode}
\hypertarget{AH8c_ad59574b55d47cec0ad8f9b18f2172608}{
\index{ad hoc.c@{ad hoc.c}!NextCard32@{NextCard32}}
\index{NextCard32@{NextCard32}!ad_hoc.c@{ad hoc.c}}
\DoxySubstituteParagraph[{NextCard32}]{\setlength{\rightskip}{0pt plus 5cm}unsigned int NextCard32 (
\begin{DoxyParamCaption}
{}
\end{DoxyParamCaption}
)}}
\label{AH8c_ad59574b55d47cec0ad8f9b18f2172608}
returns a number in the range 0...51 ( a draw of one card from a deck of 52 cards) using 32bit variables. 
\begin{DoxyCode}
\{
  \textcolor{keyword}{static} \textcolor{keywordtype}{int} l=0;
  \textcolor{keyword}{static} uint32\_t R=0;
  \textcolor{keywordflow}{if}(\hyperlink{utils_8h_ac6c45889010c1bd68631771b64f18101}{unlikely}(l<=0))\{
    R=\hyperlink{_r_n_g_8c_affb62bf72df7716e14bee3cdda966283}{my_gen_rand32}();
    l=5;
  \}
  l--;
  \textcolor{keywordtype}{unsigned} \textcolor{keywordtype}{int} c = R 
  R /= 52;
  \textcolor{keywordflow}{return} c;
\}
\end{DoxyCode}
\hypertarget{AH8c_a740a5a58d1613efb788b8d91c17845cb}{
\index{ad hoc.c@{ad hoc.c}!NextCard32 prefilled@{NextCard32 prefilled}}
\index{NextCard32 prefilled@{NextCard32 prefilled}!ad_hoc.c@{ad hoc.c}}
\DoxySubstituteParagraph[{NextCard32\textunderscore{}prefilled}]{\setlength{\rightskip}{0pt plus 5cm}unsigned int NextCard32\textunderscore{}prefilled (
\begin{DoxyParamCaption}
{}
\end{DoxyParamCaption}
)}}
\label{AH8c_a740a5a58d1613efb788b8d91c17845cb}
returns a number in the range 0...51 ( a draw of one card from a deck of 52 cards) using 32 bit variables and prefilling an array in memory. 
\begin{DoxyCode}
\{
  \textcolor{keyword}{static} \textcolor{keywordtype}{int} l=0;
  \textcolor{keywordflow}{if}(\hyperlink{utils_8h_ac6c45889010c1bd68631771b64f18101}{unlikely}(l<=0))\{
    uint32\_t R=\hyperlink{_r_n_g_8c_affb62bf72df7716e14bee3cdda966283}{my_gen_rand32}();  \textcolor{comment}{//52**5 < 2**32 }
    \_\_32saved\_cards[0] = R 
    R /= 52;
    \_\_32saved\_cards[1] = R 
    R /= 52;
    \_\_32saved\_cards[2] = R 
    R /= 52;
    \_\_32saved\_cards[3] = R 
    R /= 52;
    \_\_32saved\_cards[4] = R 
    l=5;
  \}
  l--;
  \textcolor{keywordflow}{return} \_\_32saved\_cards[l];
\}
\end{DoxyCode}
\hypertarget{AH8c_a64999255eeed624ae795239adc9cda90}{
\index{ad hoc.c@{ad hoc.c}!NextCard64@{NextCard64}}
\index{NextCard64@{NextCard64}!ad_hoc.c@{ad hoc.c}}
\DoxySubstituteParagraph[{NextCard64}]{\setlength{\rightskip}{0pt plus 5cm}unsigned int NextCard64 (
\begin{DoxyParamCaption}
{}
\end{DoxyParamCaption}
)}}
\label{AH8c_a64999255eeed624ae795239adc9cda90}
returns a number in the range 0...51 ( a draw of one card from a deck of 52 cards) using 64 bit variables. 
\begin{DoxyCode}
\{
  \textcolor{keyword}{static} \textcolor{keywordtype}{int} l=0;
  \textcolor{keyword}{static} uint64\_t R=0;
  \textcolor{keywordflow}{if}(\hyperlink{utils_8h_ac6c45889010c1bd68631771b64f18101}{unlikely}(l<=0))\{
    R=\hyperlink{_r_n_g_8c_affcfff41ea70b785738c2911275830b3}{my_gen_rand64}();
    l=11;
  \}
  l--;
  \textcolor{keywordtype}{unsigned} \textcolor{keywordtype}{int} c = R 
  R /= 52;
  \textcolor{keywordflow}{return} c;
\}
\end{DoxyCode}
\hypertarget{AH8c_aca6c7a9cd64b3924bcc52b2d1ff85ebc}{
\index{ad hoc.c@{ad hoc.c}!NextCard64 prefilled@{NextCard64 prefilled}}
\index{NextCard64 prefilled@{NextCard64 prefilled}!ad_hoc.c@{ad hoc.c}}
\DoxySubstituteParagraph[{NextCard64\textunderscore{}prefilled}]{\setlength{\rightskip}{0pt plus 5cm}unsigned int NextCard64\textunderscore{}prefilled (
\begin{DoxyParamCaption}
{}
\end{DoxyParamCaption}
)}}
\label{AH8c_aca6c7a9cd64b3924bcc52b2d1ff85ebc}
returns a number in the range 0...51 ( a draw of one card from a deck of 52 cards) using 64 bit variables and prefilling an array in memory. 
\begin{DoxyCode}
\{
  \textcolor{keyword}{static} \textcolor{keywordtype}{int} l=0;
  \textcolor{keywordflow}{if}(\hyperlink{utils_8h_ac6c45889010c1bd68631771b64f18101}{unlikely}(l<=0))\{
    uint64\_t R=\hyperlink{_r_n_g_8c_affcfff41ea70b785738c2911275830b3}{my_gen_rand64}();   \textcolor{comment}{//52**11 < 2**64}
    \_\_64saved\_cards[0] = R 
    R /= 52;
    \_\_64saved\_cards[1] = R 
    R /= 52;
    \_\_64saved\_cards[2] = R 
    R /= 52;
    \_\_64saved\_cards[3] = R 
    R /= 52;
    \_\_64saved\_cards[4] = R 
    R /= 52;
    \_\_64saved\_cards[5] = R 
    R /= 52;
    \_\_64saved\_cards[6] = R 
    R /= 52;
    \_\_64saved\_cards[7] = R 
    R /= 52;
    \_\_64saved\_cards[8] = R 
    R /= 52;
    \_\_64saved\_cards[9] = R 
    R /= 52;
    \_\_64saved\_cards[10] = R 
    l=11;
  \}
  l--;
  \textcolor{keywordflow}{return} \_\_64saved\_cards[l];
\}
\end{DoxyCode}
\hypertarget{AH8c_a1d45cfcf6adada5e5c78463dd703a4b7}{
\index{ad hoc.c@{ad hoc.c}!NextWord32@{NextWord32}}
\index{NextWord32@{NextWord32}!ad_hoc.c@{ad hoc.c}}
\DoxySubstituteParagraph[{NextWord32}]{\setlength{\rightskip}{0pt plus 5cm}unsigned int NextWord32 (
\begin{DoxyParamCaption}
{}
\end{DoxyParamCaption}
)}}
\label{AH8c_a1d45cfcf6adada5e5c78463dd703a4b7}
returns a random number in the range 0...(2$^\wedge$16-\/1), using a 32bit generator. 
\begin{DoxyCode}
\{
  \textcolor{keyword}{static} \textcolor{keywordtype}{int} l=0;
  \textcolor{keyword}{static} uint32\_t R=0;
  \textcolor{keywordflow}{if}(l<=0)\{
    R=\hyperlink{_r_n_g_8c_affb62bf72df7716e14bee3cdda966283}{my_gen_rand32}();
    l=2;
  \}
  \textcolor{keywordtype}{unsigned} \textcolor{keywordtype}{int} bytes=R & 0xFFFF;
  R>>=16;
  l--;
  \textcolor{keywordflow}{return} bytes;
\}
\end{DoxyCode}
\hypertarget{AH8c_a9c0eee176aba60fd9c925fed0b93722a}{
\index{ad hoc.c@{ad hoc.c}!NextWord64@{NextWord64}}
\index{NextWord64@{NextWord64}!ad_hoc.c@{ad hoc.c}}
\DoxySubstituteParagraph[{NextWord64}]{\setlength{\rightskip}{0pt plus 5cm}unsigned int NextWord64 (
\begin{DoxyParamCaption}
{}
\end{DoxyParamCaption}
)}}
\label{AH8c_a9c0eee176aba60fd9c925fed0b93722a}
returns a random number in the range 0...(2$^\wedge$16-\/1), using a 64bit generator. 
\begin{DoxyCode}
\{
  \textcolor{keyword}{static} \textcolor{keywordtype}{int} l=0;
  \textcolor{keyword}{static} uint64\_t R=0;
  \textcolor{keywordflow}{if}(\hyperlink{utils_8h_ac6c45889010c1bd68631771b64f18101}{unlikely}(l<=0))\{
    R=\hyperlink{_r_n_g_8c_affcfff41ea70b785738c2911275830b3}{my_gen_rand64}();
    l=4;
  \}
  \textcolor{keywordtype}{unsigned} \textcolor{keywordtype}{int} bytes=R & 0xFFFF;
  R>>=16;
  l--;
  \textcolor{keywordflow}{return} bytes;
\}
\end{DoxyCode}

}


\subsection{Uniform random functions}
\label{sec:ur_C}

{\small
\hypertarget{urf8c}{
\subsubsection{uniform random func.c File Reference}
\label{urf8c}\index{uniform random func.c@{uniform random func.c}}
}

Uniform random functions tested in the paper.

\subsubsection*{Defines}
\begin{DoxyCompactItemize}
\item 
\#define \hyperlink{urf8c_aa5a0f328018cb09f3af98f640cc68960}{COUNTFAILURES}(A)
\end{DoxyCompactItemize}
\subsubsection*{Functions}
\begin{DoxyCompactItemize}
\item 
uint32\textunderscore{}t \hyperlink{urf8c_a6af7ab40e5ca842b417ac842d2d288da}{uniform\textunderscore{}random\textunderscore{}by\textunderscore{}bit\textunderscore{}recycling} (uint32\textunderscore{}t n)
\item 
uint32\textunderscore{}t \hyperlink{urf8c_ad8fa334e03dbc59cfcf0aa2b0cf48730}{uniform\textunderscore{}random\textunderscore{}by\textunderscore{}bit\textunderscore{}recycling\textunderscore{}faster} (uint32\textunderscore{}t n)
\item 
uint32\textunderscore{}t \hyperlink{urf8c_ad7595115a21ecb206863cd0cde74ecbf}{uniform\textunderscore{}random\textunderscore{}by\textunderscore{}bit\textunderscore{}recycling\textunderscore{}cheating} (uint32\textunderscore{}t n)
\item 
uint32\textunderscore{}t \hyperlink{urf8c_a0a08b4b7cbfb41c660f50126582f43b7}{uniform\textunderscore{}random\textunderscore{}by\textunderscore{}bit\textunderscore{}recycling32} (uint32\textunderscore{}t n)
\item 
uint32\textunderscore{}t \hyperlink{urf8c_a6d5c69dae88fbd8fd093f13adddbb27f}{uniform\textunderscore{}random\textunderscore{}simple\textunderscore{}64} (uint32\textunderscore{}t n)
\item 
uint32\textunderscore{}t \hyperlink{urf8c_a8ee1e8d21b0e86cf383bb532cb4b7d4c}{uniform\textunderscore{}random\textunderscore{}simple\textunderscore{}32} (uint32\textunderscore{}t n)
\item 
uint32\textunderscore{}t \hyperlink{urf8c_acde4feae3f48f9b97b92ec56983cf259}{uniform\textunderscore{}random\textunderscore{}simple\textunderscore{}recycler} (uint32\textunderscore{}t n)
\item 
uint32\textunderscore{}t \hyperlink{urf8c_aa7838125a9d0dbaeb62c030ec5d32d97}{uniform\textunderscore{}random\textunderscore{}simple\textunderscore{}40} (uint32\textunderscore{}t n)
\item 
uint32\textunderscore{}t \hyperlink{urf8c_aa5bf6735d2d319711e0d99fb25cd98ad}{uniform\textunderscore{}random\textunderscore{}simple\textunderscore{}48} (uint32\textunderscore{}t n)
\end{DoxyCompactItemize}

\subsubsection{Detailed Description}
In all the functions, the argument 'n' is the modulus of the returned uniform random number. 

\subsubsection{Define Documentation}
\hypertarget{urf8c_aa5a0f328018cb09f3af98f640cc68960}{
\index{uniform random func.c@{uniform random func.c}!COUNTFAILURES@{COUNTFAILURES}}
\index{COUNTFAILURES@{COUNTFAILURES}!uniform_random_func.c@{uniform random func.c}}
\DoxySubstituteParagraph[{COUNTFAILURES}]{\setlength{\rightskip}{0pt plus 5cm}\#define COUNTFAILURES(
\begin{DoxyParamCaption}
\item[{}]{A}
\end{DoxyParamCaption}
)}}
\label{urf8c_aa5a0f328018cb09f3af98f640cc68960}
COUNTFAILURE is defined either as

\begin{DoxyItemize}
\item \#define \hyperlink{urf8c_aa5a0f328018cb09f3af98f640cc68960}{COUNTFAILURES(A)}
\end{DoxyItemize}

or as

\begin{DoxyItemize}
\item \#define \hyperlink{urf8c_aa5a0f328018cb09f3af98f640cc68960}{COUNTFAILURES(A)} global\textunderscore{}counter++;
\end{DoxyItemize}

depending on if we want to count failures 

\subsubsection{Function Documentation}
\hypertarget{urf8c_a6af7ab40e5ca842b417ac842d2d288da}{
\index{uniform random func.c@{uniform random func.c}!uniform random by bit recycling@{uniform random by bit recycling}}
\index{uniform random by bit recycling@{uniform random by bit recycling}!uniform_random_func.c@{uniform random func.c}}
\DoxySubstituteParagraph[{uniform\textunderscore{}random\textunderscore{}by\textunderscore{}bit\textunderscore{}recycling}]{\setlength{\rightskip}{0pt plus 5cm}uint32\textunderscore{}t uniform\textunderscore{}random\textunderscore{}by\textunderscore{}bit\textunderscore{}recycling (
\begin{DoxyParamCaption}
\item[{uint32\textunderscore{}t}]{n}
\end{DoxyParamCaption}
)}}
\label{urf8c_a6af7ab40e5ca842b417ac842d2d288da}
This implements the pseudocode in figure 2 (but pops 2 bits at a time). 
\begin{DoxyCode}
\{
  \textcolor{keyword}{static} uint64\_t m = 1, r = 0;
  \textcolor{keyword}{const} uint64\_t N62=((uint64\_t)1)<<62;
  \textcolor{keywordflow}{while}(1)\{
    \textcolor{keywordflow}{while}(m<N62)\{ \textcolor{comment}{//fill the state}
      r = ( r << 2 ) | \hyperlink{optimizer__sfmt__sse__fixed_8h_a0c1cc90d5f223a5ec134000d1f5e7a58}{Next2Bit}();
      m =   m << 2;
    \}
    \textcolor{keyword}{const} uint64\_t q=m/n, nq = n * q;
    \textcolor{keywordflow}{if}( \hyperlink{utils_8h_a217a0bd562b98ae8c2ffce44935351e1}{likely}(r < nq) )\{
      uint32\_t d = r 
      r = r / n;            \textcolor{comment}{//quotient, is random variable of modulus q}
      m = q;
      \textcolor{keywordflow}{return}  d;
    \} \textcolor{keywordflow}{else}\{
      \hyperlink{urf8c_aa5a0f328018cb09f3af98f640cc68960}{COUNTFAILURES}();
      m = m - nq;
      r = r - nq; \textcolor{comment}{// r is still a random variable of modulus m}
    \}
  \}
\}
\end{DoxyCode}
\hypertarget{urf8c_a0a08b4b7cbfb41c660f50126582f43b7}{
\index{uniform random func.c@{uniform random func.c}!uniform random by bit recycling32@{uniform random by bit recycling32}}
\index{uniform random by bit recycling32@{uniform random by bit recycling32}!uniform_random_func.c@{uniform random func.c}}
\DoxySubstituteParagraph[{uniform\textunderscore{}random\textunderscore{}by\textunderscore{}bit\textunderscore{}recycling32}]{\setlength{\rightskip}{0pt plus 5cm}uint32\textunderscore{}t uniform\textunderscore{}random\textunderscore{}by\textunderscore{}bit\textunderscore{}recycling32 (
\begin{DoxyParamCaption}
\item[{uint32\textunderscore{}t}]{n}
\end{DoxyParamCaption}
)}}
\label{urf8c_a0a08b4b7cbfb41c660f50126582f43b7}
This implements the pseudocode in figure 2 but only using 32bit variables, so it has some special methods when n$>$=2$^\wedge$29. 
\begin{DoxyCode}
\{
  \textcolor{keyword}{const} uint32\_t N29=((uint32\_t)1)<<29,  N30=((uint32\_t)1)<<30,  N31=((uint32\_t)1
      )<<31, N24=((uint32\_t)1)<<24;
  \textcolor{comment}{//special methods}
  \textcolor{keywordflow}{if} ( n >= N31 )\{
    \textcolor{keywordflow}{while}(1)\{
      uint32\_t r = \hyperlink{_r_n_g_8c_affb62bf72df7716e14bee3cdda966283}{my_gen_rand32}();
      \textcolor{keywordflow}{if}( \hyperlink{utils_8h_a217a0bd562b98ae8c2ffce44935351e1}{likely}(r < n) )\{
        \textcolor{keywordflow}{return}  r;
      \} \textcolor{keywordflow}{else}\{
        \hyperlink{urf8c_aa5a0f328018cb09f3af98f640cc68960}{COUNTFAILURES}();
      \}
    \}\}
  \textcolor{keywordflow}{if} ( n >= N30 )\{
    \textcolor{keywordflow}{while}(1)\{
      uint32\_t r = \hyperlink{_r_n_g_8c_affb62bf72df7716e14bee3cdda966283}{my_gen_rand32}() >> 1;
      \textcolor{keywordflow}{if}( \hyperlink{utils_8h_a217a0bd562b98ae8c2ffce44935351e1}{likely}(r < n) )\{
        \textcolor{keywordflow}{return}  r;
      \} \textcolor{keywordflow}{else}\{
        \hyperlink{urf8c_aa5a0f328018cb09f3af98f640cc68960}{COUNTFAILURES}();
      \}
    \}\}
  \textcolor{keywordflow}{if} ( n >= N29 )\{
    \textcolor{keywordflow}{while}(1)\{
      uint32\_t r = \hyperlink{_r_n_g_8c_affb62bf72df7716e14bee3cdda966283}{my_gen_rand32}() >> 2;
      \textcolor{keywordflow}{if}( \hyperlink{utils_8h_a217a0bd562b98ae8c2ffce44935351e1}{likely}(r < n) )\{
        \textcolor{keywordflow}{return}  r;
      \} \textcolor{keywordflow}{else}\{
        \hyperlink{urf8c_aa5a0f328018cb09f3af98f640cc68960}{COUNTFAILURES}();
      \}
    \}\}
  \textcolor{comment}{//usual bit recycling}
  \textcolor{keyword}{static} uint32\_t m = 1, r = 0;
  \textcolor{keywordflow}{while}(1)\{
    \textcolor{keywordflow}{while} ( m < N24 )\{ \textcolor{comment}{//fill the state}
      r = ( r << 8 ) | \hyperlink{optimizer__sfmt__sse__fixed_8h_af2d3ba63dc33f940f210ecb9d2cc1f3f}{NextByte}();
      m =   m << 8;
    \}
    \textcolor{keywordflow}{while}(m<N30)\{ \textcolor{comment}{//fill the state}
      r = ( r << 2 ) | \hyperlink{optimizer__sfmt__sse__fixed_8h_a0c1cc90d5f223a5ec134000d1f5e7a58}{Next2Bit}();
      m =   m << 2;
    \}
    uint32\_t q = m / n, nq = q * n;
    \textcolor{keywordflow}{if}(\hyperlink{utils_8h_a217a0bd562b98ae8c2ffce44935351e1}{likely}( r < nq) )\{
      uint32\_t d = r 
      r = r / n;            \textcolor{comment}{//quotient, is random variable of modulus q}
      m = q;
      \textcolor{keywordflow}{return} d;
    \} \textcolor{keywordflow}{else}\{
      \hyperlink{urf8c_aa5a0f328018cb09f3af98f640cc68960}{COUNTFAILURES}();
      m = m - nq;
      r = r - nq; \textcolor{comment}{// r is still a random variable of modulus m}
    \}
  \}
\}
\end{DoxyCode}
\hypertarget{urf8c_ad7595115a21ecb206863cd0cde74ecbf}{
\index{uniform random func.c@{uniform random func.c}!uniform random by bit recycling cheating@{uniform random by bit recycling cheating}}
\index{uniform random by bit recycling cheating@{uniform random by bit recycling cheating}!uniform_random_func.c@{uniform random func.c}}
\DoxySubstituteParagraph[{uniform\textunderscore{}random\textunderscore{}by\textunderscore{}bit\textunderscore{}recycling\textunderscore{}cheating}]{\setlength{\rightskip}{0pt plus 5cm}uint32\textunderscore{}t uniform\textunderscore{}random\textunderscore{}by\textunderscore{}bit\textunderscore{}recycling\textunderscore{}cheating (
\begin{DoxyParamCaption}
\item[{uint32\textunderscore{}t}]{n}
\end{DoxyParamCaption}
)}}
\label{urf8c_ad7595115a21ecb206863cd0cde74ecbf}
This implements the pseudocode in figure 2, but does not implement the \char`\"{}else\char`\"{} block, so it is not mathematically perfect; at the same time, since the probability of the \char`\"{}else\char`\"{} block would be less than 1/2$^\wedge$24 the random numbers generated by this function are good enough for most purposes. 
\begin{DoxyCode}
\{
  \textcolor{keyword}{static} uint64\_t m = 1, r = 0;
  \textcolor{keyword}{const} uint64\_t
    N48=((uint64\_t)1)<<48, N56=((uint64\_t)1)<<56;
  \textcolor{keywordflow}{if}(m<N48)\{   \textcolor{comment}{//fill the state}
    r = ( r << 16 ) | \hyperlink{optimizer__sfmt__sse__fixed_8h_add93353118e776ad57d8ea393a7353f3}{NextWord}();
    m =   m << 16;
  \}
  \textcolor{keywordflow}{while}(m<N56)\{
    r = ( r << 8 ) | \hyperlink{optimizer__sfmt__sse__fixed_8h_af2d3ba63dc33f940f210ecb9d2cc1f3f}{NextByte}();
    m =   m << 8;
  \}
  uint32\_t d = r 
  r = r / n;
  m = m / n;
  \textcolor{keywordflow}{return}  d;
\}
\end{DoxyCode}
\hypertarget{urf8c_ad8fa334e03dbc59cfcf0aa2b0cf48730}{
\index{uniform random func.c@{uniform random func.c}!uniform random by bit recycling faster@{uniform random by bit recycling faster}}
\index{uniform random by bit recycling faster@{uniform random by bit recycling faster}!uniform_random_func.c@{uniform random func.c}}
\DoxySubstituteParagraph[{uniform\textunderscore{}random\textunderscore{}by\textunderscore{}bit\textunderscore{}recycling\textunderscore{}faster}]{\setlength{\rightskip}{0pt plus 5cm}uint32\textunderscore{}t uniform\textunderscore{}random\textunderscore{}by\textunderscore{}bit\textunderscore{}recycling\textunderscore{}faster (
\begin{DoxyParamCaption}
\item[{uint32\textunderscore{}t}]{n}
\end{DoxyParamCaption}
)}}
\label{urf8c_ad8fa334e03dbc59cfcf0aa2b0cf48730}
This implements the pseudocode in figure 2, but pops words,bytes,and pairs of bits. 
\begin{DoxyCode}
\{
  \textcolor{keyword}{static} uint64\_t m = 1, r = 0;
  \textcolor{keyword}{const} uint64\_t
    N62 = ((uint64\_t)1) << 62,
    N56 = ((uint64\_t)1) << 56,
    N48 = ((uint64\_t)1) << 48;
  \textcolor{keywordflow}{while}(1)\{
    \textcolor{comment}{//fill the state}
    \textcolor{keywordflow}{if} ( m < N48 )\{
      r = ( r << 16 ) | \hyperlink{optimizer__sfmt__sse__fixed_8h_add93353118e776ad57d8ea393a7353f3}{NextWord}();
      m =   m << 16;
    \}
    \textcolor{keywordflow}{while} ( m < N56 )\{
      r = ( r << 8 ) | \hyperlink{optimizer__sfmt__sse__fixed_8h_af2d3ba63dc33f940f210ecb9d2cc1f3f}{NextByte}();
      m =   m << 8;
    \}
    \textcolor{keywordflow}{while} ( m < N62)\{
      r = ( r << 2 ) | \hyperlink{optimizer__sfmt__sse__fixed_8h_a0c1cc90d5f223a5ec134000d1f5e7a58}{Next2Bit}();
      m =   m << 2;
    \}
    \textcolor{keyword}{const} uint64\_t q=m/n, nq=n*q;
    \textcolor{keywordflow}{if}( \hyperlink{utils_8h_a217a0bd562b98ae8c2ffce44935351e1}{likely}(r < nq) )\{
      uint32\_t d = r 
      r = r / n;          \textcolor{comment}{//quotient, is random variable of modulus q}
      m = q;
      \textcolor{keywordflow}{return}  d;
    \} \textcolor{keywordflow}{else}\{
      \hyperlink{urf8c_aa5a0f328018cb09f3af98f640cc68960}{COUNTFAILURES}();
      m = m - nq;
      r = r - nq; \textcolor{comment}{// r is still a random variable of modulus m}
    \}
  \}
\}
\end{DoxyCode}
\hypertarget{urf8c_a8ee1e8d21b0e86cf383bb532cb4b7d4c}{
\index{uniform random func.c@{uniform random func.c}!uniform random simple 32@{uniform random simple 32}}
\index{uniform random simple 32@{uniform random simple 32}!uniform_random_func.c@{uniform random func.c}}
\DoxySubstituteParagraph[{uniform\textunderscore{}random\textunderscore{}simple\textunderscore{}32}]{\setlength{\rightskip}{0pt plus 5cm}uint32\textunderscore{}t uniform\textunderscore{}random\textunderscore{}simple\textunderscore{}32 (
\begin{DoxyParamCaption}
\item[{uint32\textunderscore{}t}]{n}
\end{DoxyParamCaption}
)}}
\label{urf8c_a8ee1e8d21b0e86cf383bb532cb4b7d4c}
This is a simple implementation, found in many random number libraries; this version uses 32 bit variables. 
\begin{DoxyCode}
\{
  \textcolor{keyword}{const} uint32\_t N = 0xFFFFFFFFU; \textcolor{comment}{// 2^32-1;}
  uint32\_t q = N / n;
  \textcolor{keywordflow}{while}(1)\{
    uint32\_t r = \hyperlink{_r_n_g_8c_affb62bf72df7716e14bee3cdda966283}{my_gen_rand32}();
    \textcolor{keywordflow}{if}( \hyperlink{utils_8h_a217a0bd562b98ae8c2ffce44935351e1}{likely}(r < n * q))\{
      uint32\_t d = r 
      \textcolor{keywordflow}{return}  d;
    \} \textcolor{keywordflow}{else}\{
      \hyperlink{urf8c_aa5a0f328018cb09f3af98f640cc68960}{COUNTFAILURES}();
    \}
  \}
\}
\end{DoxyCode}
\hypertarget{urf8c_aa7838125a9d0dbaeb62c030ec5d32d97}{
\index{uniform random func.c@{uniform random func.c}!uniform random simple 40@{uniform random simple 40}}
\index{uniform random simple 40@{uniform random simple 40}!uniform_random_func.c@{uniform random func.c}}
\DoxySubstituteParagraph[{uniform\textunderscore{}random\textunderscore{}simple\textunderscore{}40}]{\setlength{\rightskip}{0pt plus 5cm}uint32\textunderscore{}t uniform\textunderscore{}random\textunderscore{}simple\textunderscore{}40 (
\begin{DoxyParamCaption}
\item[{uint32\textunderscore{}t}]{n}
\end{DoxyParamCaption}
)}}
\label{urf8c_aa7838125a9d0dbaeb62c030ec5d32d97}
some alternative versions, using 64 bit variables and 40bit state 
\begin{DoxyCode}
\{
  \textcolor{keyword}{const} uint64\_t N = ((uint64\_t)1) << 40;
  uint64\_t q = N / n, nq = ((uint64\_t)n) * q;
  \textcolor{keywordflow}{while}(1)\{
    uint64\_t r = \hyperlink{_r_n_g_8c_affb62bf72df7716e14bee3cdda966283}{my_gen_rand32}();
    r=(r<<8) | \hyperlink{optimizer__sfmt__sse__fixed_8h_af2d3ba63dc33f940f210ecb9d2cc1f3f}{NextByte}(); \textcolor{comment}{//create 40 bits random numbers}
    \textcolor{keywordflow}{if}( \hyperlink{utils_8h_a217a0bd562b98ae8c2ffce44935351e1}{likely}(r < nq) )\{
      uint32\_t d = r 
      \textcolor{keywordflow}{return}  d;
    \} \textcolor{keywordflow}{else}\{
      \hyperlink{urf8c_aa5a0f328018cb09f3af98f640cc68960}{COUNTFAILURES}();
    \}
  \}
\}
\end{DoxyCode}
\hypertarget{urf8c_aa5bf6735d2d319711e0d99fb25cd98ad}{
\index{uniform random func.c@{uniform random func.c}!uniform random simple 48@{uniform random simple 48}}
\index{uniform random simple 48@{uniform random simple 48}!uniform_random_func.c@{uniform random func.c}}
\DoxySubstituteParagraph[{uniform\textunderscore{}random\textunderscore{}simple\textunderscore{}48}]{\setlength{\rightskip}{0pt plus 5cm}uint32\textunderscore{}t uniform\textunderscore{}random\textunderscore{}simple\textunderscore{}48 (
\begin{DoxyParamCaption}
\item[{uint32\textunderscore{}t}]{n}
\end{DoxyParamCaption}
)}}
\label{urf8c_aa5bf6735d2d319711e0d99fb25cd98ad}
some alternative versions, using 64 bit variables and 48bit state 
\begin{DoxyCode}
\{
  \textcolor{keyword}{const} uint64\_t N=((uint64\_t)1)<<48;
  uint64\_t q = N/n, nq=((uint64\_t)n) * q;
  \textcolor{keywordflow}{while}(1)\{
    uint64\_t r = \hyperlink{_r_n_g_8c_affb62bf72df7716e14bee3cdda966283}{my_gen_rand32}();
    r=(r<<16) | \hyperlink{optimizer__sfmt__sse__fixed_8h_add93353118e776ad57d8ea393a7353f3}{NextWord}(); \textcolor{comment}{//create 48 bits random numbers}
    \textcolor{keywordflow}{if}( \hyperlink{utils_8h_a217a0bd562b98ae8c2ffce44935351e1}{likely}(r < nq) )\{
      uint32\_t d = r 
      \textcolor{keywordflow}{return}  d;
    \} \textcolor{keywordflow}{else}\{
      \hyperlink{urf8c_aa5a0f328018cb09f3af98f640cc68960}{COUNTFAILURES}();
    \}
  \}
\}
\end{DoxyCode}
\hypertarget{urf8c_a6d5c69dae88fbd8fd093f13adddbb27f}{
\index{uniform random func.c@{uniform random func.c}!uniform random simple 64@{uniform random simple 64}}
\index{uniform random simple 64@{uniform random simple 64}!uniform_random_func.c@{uniform random func.c}}
\DoxySubstituteParagraph[{uniform\textunderscore{}random\textunderscore{}simple\textunderscore{}64}]{\setlength{\rightskip}{0pt plus 5cm}uint32\textunderscore{}t uniform\textunderscore{}random\textunderscore{}simple\textunderscore{}64 (
\begin{DoxyParamCaption}
\item[{uint32\textunderscore{}t}]{n}
\end{DoxyParamCaption}
)}}
\label{urf8c_a6d5c69dae88fbd8fd093f13adddbb27f}
This is a simple implementation, found in many random number libraries; this version uses 64 bit variables. 
\begin{DoxyCode}
\{
  \textcolor{keyword}{const} uint64\_t N = 0xFFFFFFFFFFFFFFFFU; \textcolor{comment}{//2^64-1;}
  uint64\_t q = N / n, nq = ((uint64\_t)n) * q;
  \textcolor{keywordflow}{while}(1)\{
    uint64\_t r = \hyperlink{_r_n_g_8c_affcfff41ea70b785738c2911275830b3}{my_gen_rand64}();
    \textcolor{keywordflow}{if}( \hyperlink{utils_8h_a217a0bd562b98ae8c2ffce44935351e1}{likely}(r < nq) )\{
      uint32\_t d = r 
      \textcolor{keywordflow}{return}  d;
    \} \textcolor{keywordflow}{else}\{
      \hyperlink{urf8c_aa5a0f328018cb09f3af98f640cc68960}{COUNTFAILURES}();
    \}
  \}
\}
\end{DoxyCode}
\hypertarget{urf8c_acde4feae3f48f9b97b92ec56983cf259}{
\index{uniform random func.c@{uniform random func.c}!uniform random simple recycler@{uniform random simple recycler}}
\index{uniform random simple recycler@{uniform random simple recycler}!uniform_random_func.c@{uniform random func.c}}
\DoxySubstituteParagraph[{uniform\textunderscore{}random\textunderscore{}simple\textunderscore{}recycler}]{\setlength{\rightskip}{0pt plus 5cm}uint32\textunderscore{}t uniform\textunderscore{}random\textunderscore{}simple\textunderscore{}recycler (
\begin{DoxyParamCaption}
\item[{uint32\textunderscore{}t}]{n}
\end{DoxyParamCaption}
)}}
\label{urf8c_acde4feae3f48f9b97b92ec56983cf259}
This is a simple implementation, with recycling for small n this version uses 32 bit variables. 
\begin{DoxyCode}
\{
  \textcolor{keyword}{static}  uint32\_t \_r = 0, \_m = 0;
  \textcolor{keyword}{const} uint32\_t N = 0xFFFFFFFFU; \textcolor{comment}{//2^32-1}
  \textcolor{keywordflow}{if} (\_m>n)\{
      uint32\_t d = \_r 
      \_m /= n;
      \_r /= n;
      \textcolor{keywordflow}{return} d;
    \}
  \textcolor{keyword}{const} uint32\_t q = N / n, nq = n * q;
  \textcolor{keywordflow}{while}(1)\{
    uint32\_t newr = \hyperlink{_r_n_g_8c_affb62bf72df7716e14bee3cdda966283}{my_gen_rand32}();
    \textcolor{keywordflow}{if}(newr < nq)\{
      uint32\_t d = newr 
      \textcolor{keywordflow}{if}(\_m < q )\{ \textcolor{comment}{//there is more entropy in newr than in \_r}
        \_m = q;
        \_r = newr / n;
      \}
      \textcolor{keywordflow}{return}  d;
    \} \textcolor{keywordflow}{else}\{
      \hyperlink{urf8c_aa5a0f328018cb09f3af98f640cc68960}{COUNTFAILURES}();
    \}
  \}
\}
\end{DoxyCode}

}


\section*{Acknowledgments}
The author thanks Professors S. Marmi and A. Profeti
for allowing access to hardware.

\tableofcontents




\begin{thebibliography}{10}

\bibitem[1]{DJ}
``Doctor Jacques'' (2004) at the Math forum
\newblock \url{http://mathforum.org/library/drmath/view/65653.html}

\bibitem[2]{SFMT}
 Mutsuo Saito and Makoto Matsumoto (2008),
\newblock ``SIMD oriented Fast Mersenne Twister(SFMT): a 128-bit Pseudorandom Number Generator.''
\newblock ``Monte Carlo and Quasi-Monte Carlo Methods 2006'', Springer, pp. 607 -- 622.
\newblock DOI \href{http://dx.doi.org/10.1007/978-3-540-74496-2_36}{10.1007/978-3-540-74496-2\_36}.
\newblock \textsl{The source code  (ver. 1.3.3) is available from}
\url{http://www.math.sci.hiroshima-u.ac.jp/~m-mat/MT/SFMT/index.html}

\bibitem[3]{dSFMT} 
 Mutsuo Saito and Makoto Matsumoto (2009),
\newblock ``A PRNG Specialized in Double Precision Floating Point Number Using an Affine Transition.''
\newblock ``Monte Carlo and Quasi-Monte Carlo Methods 2008'', Springer, pp. 589 -- 602. 
\newblock DOI
\href{http://dx.doi.org/10.1007/978-3-642-04107-5_38}{10.1007/978-3-642-04107-5\_38}


\bibitem[4]{xorshift}
George Marsaglia (2003)
\newblock ``Xorshift RNGs.''
\newblock Journal of Statistical Software, 8, 1-9. 
\newblock \url{http://www.jstatsoft.org/v08/i14/}.


\bibitem[5]{bbs}
L. Blum, M. Blum, and M. Shub (1986), 
\newblock ``A Simple Unpredictable Pseudo-Random Number Generator.''
\newblock \emph{SIAM J. Comput. 15, 364},
\newblock DOI \href{http://dx.doi.org/10.1137/0215025}{10.1137/0215025}


\end{thebibliography}
\end{document}